\documentstyle[12pt,aaspp4]{article}
\makeatletter

\@ifundefined{chapter}{\def\thebibliography#1{\section*{References\@mkboth
  {REFERENCES}{REFERENCES}}\list
  {\relax}{\setlength{\labelsep}{0em}
        \setlength{\itemindent}{-\bibhang}
        \setlength{\itemsep}{0pt}
        \setlength{\parsep}{0pt}
        \setlength{\leftmargin}{\bibhang}}
    \def\newblock{\hskip .11em plus .33em minus .07em}
    \sloppy\clubpenalty4000\widowpenalty4000
    \sfcode`\.=1000\relax}}%
{\def\thebibliography#1{\chapter*{Bibliography\@mkboth
  {BIBLIOGRAPHY}{BIBLIOGRAPHY}}\list
  {\relax}{\setlength{\labelsep}{0em}
        \setlength{\itemindent}{-\bibhang}
        \setlength{\itemsep}{0pt}
        \setlength{\parsep}{0pt}
        \setlength{\leftmargin}{\bibhang}}
    \def\newblock{\hskip .11em plus .33em minus .07em}
    \sloppy\clubpenalty4000\widowpenalty4000
    \sfcode`\.=1000\relax}}

\newlength{\bibhang}
\let\@internalcite\cite
\def\cite{\let\@citeleft(\let\@citeright)%
    \@ifstar{\citeyear}{\citefull}}
\def\acite{\let\@citeleft\relax\let\@citeright\relax%
    \@ifstar{\citeyear}{\acitefull}}
\def\citenp{\let\@citeleft\relax\let\@citeright\relax
    \@ifstar{\citeyear}{\citefull}}
\def\citefull{\def\astroncite##1##2{##1~##2}\@internalcite}
\def\citeyear{\def\astroncite##1##2{##2}\@internalcite}
\def\acitefull{\def\astroncite##1##2{##1~(##2)}\@internalcite}

\def\@citex[#1]#2{\if@filesw\immediate\write\@auxout{\string\citation{#2}}\fi
  \def\@citea{}\@cite{\@for\@citeb:=#2\do
    {\@citea\def\@citea{; }\@ifundefined
       {b@\@citeb}{{\bf ?}\@warning
       {Citation `\@citeb' on page \thepage \space undefined}}%
{\csname b@\@citeb\endcsname}}}{#1}}

\def\@cite#1#2{\@citeleft#1\if@tempswa , #2\fi\@citeright}
\def\@biblabel#1{}

\makeatother

\newcommand{\PSbox}[3]{\mbox{\rule{0in}{#3}\includegraphics{#1}\hspace{#2}}}
\newcommand{\FigNum}[1]{\unitlength 1pt \begin{picture}(55,10)(-400,35) 
                        \put(0,0){Figure #1}
                        \end{picture}}

\newcommand{\persec}{\mbox{$\second^{-1}$}}
\newcommand{\percm}{\mbox{$\cm^{-2}$}}
\newcommand{\ppm}{\mbox{$\pm$}}
\newcommand{\cgsflux}{\erg\percm\persec}
\newcommand{\cgslum}{\erg\persec}

\newcommand\approxgt{\mbox{$^{>}\hspace{-0.24cm}_{\sim}$}}
\newcommand\approxlt{\mbox{$^{<}\hspace{-0.24cm}_{\sim}$}}
\def\etal{{et~al.}}

\newcommand{\nh}{\mbox{$N_{\rm H}$}}

\newcommand{\ee}[1]{\mbox{$10^{#1}$}}
\newcommand{\tee}[1]{\mbox{$\times 10^{#1}$}}

\newcommand{\sig}{{$\sigma$}}
\newcommand\lxlbol{$L_{X}$/$L_{\rm bol}$}  
\newcommand\lxlopt{\mbox{$L_{X}/L_{\rm opt}$}}

\newcommand{\perval}[2]{{#1\mbox{$^{#2}$}}}

\def\x1608{{4U~1608$-$522}}

\def\cenx4{{Cen~X$-$4}}

\def\saxj1808{{SAX J1808.4$-$3658}}

\newcommand{\cm}{\mbox{$\rm\,cm$}}

\newcommand{\second}{\mbox{$\rm\,s$}}

\newcommand{\erg}{\mbox{$\rm\,erg$}}

\newcommand{\kteff}{\mbox{$kT_{\rm eff}$}}

\newcommand{\chandra}{{\em Chandra\/}}

\newcommand{\rosat}{{\em ROSAT\/}}

\newcommand{\usno}{{USNO-A2}}
\newcommand{\pid}{\mbox{$P_{\rm id}$}}
\newcommand{\pnoid}{\mbox{$P_{\rm no-id}$}}

\begin{document}

\title{A Limit on the Number of Isolated Neutron Stars Detected in the
ROSAT Bright Source Catalog}

\author{Robert E. Rutledge, Derek W. Fox, Milan Bogosavljevic, Ashish Mahabal\altaffilmark{1}} \altaffiltext{1}{ California Institute of
Technology, MS 130-33, Pasadena, CA 91125; rutledge@tapir.caltech.edu,
derekfox@astro.caltech.edu, milan@astro.caltech.edu, aam@astro.caltech.edu}

\begin{abstract}
The challenge in searching for non-radio-pulsing isolated neutron
stars (INSs) is in excluding association with objects in the very
large error boxes ($\sim$13\arcsec, 1$\sigma$ radius) typical of
sources from the largest X-ray all-sky survey, the \rosat\
All-Sky-Survey/Bright Source Catalog (RASS/BSC).  We search for
candidate INSs using statistical analysis of optical (USNO-A2),
infrared (IRAS), and radio (NVSS) sources near the \rosat\ X-ray
localization, and show that this selection would find 20\% of the INSs
in the RASS/BSC.  This selection finds 32 candidates at
declinations $\delta>-39\, \deg$, among which are two previously known
INSs, seventeen sources which we show are not INSs, and thirteen the
classification of which are as yet undetermined.  These results
require a limit of $<$67 INSs (90\% confidence, full sky, assuming
isotropy) in the RASS/BSC.  This limit modestly constrains a
naive and optimistic model for cooling NSs in the galaxy. 
\end{abstract}

\section{Introduction}

Initial estimates of the number of isolated neutron stars (INSs),
accreting through the Bondi (gravitational) mode
\cite{bondi44,bondi52}, which would be detected in the \rosat All-Sky
Survey (RASS) were of the order \ee{3}--\ee{4}
\cite{treves91,blaes93,madau94}.  As observations of higher mean
velocities in the radio pulsar population \cite[and references
therein]{lorimer97,hansen97,cordes98} were considered, the predicted
number of detectable INSs accreting from the ISM decreased
dramatically ($\sim$\ee{2}--\ee{3};
\citenp{colpi98,neuhauser99b,treves00}).  This is because the Bondi
accretion rate is a strong function of the NS velocity through the
interstellar medium (ISM; $\dot{M}\propto v^{-3}$).

Considerable efforts to discover INSs have been applied.  Many have
used the ROSAT/All-Sky Survey Bright Source Catalog (RASS/BSC;
\citenp{voges99}), which contains the 18,811 brightest X-ray sources
detected in a survey with the \rosat/PSPC in 1990/1991, with
positional certainties of $\sim$12\arcsec\ (1$\sigma$).  The survey
covers 92\% of the sky in the 0.1-2.4 keV band; the RASS sources are
complete down to a PSPC countrate of 0.05 c/s, which corresponds to a
flux of 5.4\tee{-13} \cgsflux\ assuming an unabsorbed power-law
spectrum of photon index $\alpha=1$ (or a flux of 2\tee{-13} assuming
$\alpha=3$).  These efforts searched for error boxes which are
``empty'' of off-band counterparts. INSs have an X-ray to optical
ratio of $\sim$\ee{5}, substantially greater than any other class of
X-ray sources.  The X-ray to optical ratio of a thermal spectrum INS
is roughly \cite{treves00}:
\begin{equation}
\frac{L_X}{L_{\rm opt}} \sim 10 ^{5.5 + 3 \log \left( \frac{kT_{\rm eff}}{100 {\rm eV}} \right)}
\end{equation}

\noindent which is orders of magnitude larger than for other known
X-ray source classes; for example, stars typically have
\lxlbol\approxlt\ee{-3}, AGN are $\sim$0.1-10, and white dwarfs and
X-ray binaries are typically 10--100, but can be as high as 1000.  To
date, three bright INSs have been optically detected near their 
expected flux level \cite{walter97,kulkarni98,kaplan02}.  No
identified INS has exhibited intensity variability on any timescale. 

The predictions for the number of INSs to be detected contrast with
only seven such objects reported so far \cite{treves00}.  This number
is so low, that it has been suggested that these INSs are powered not
by accretion, but by hot cores of young, cooling NSs
\cite{neuhauser99b,popov00b}.  We argue in \S~\ref{sec:confusion} that
previous analyses have not quantitatively justified their source
confusion rates, and the resultant limits on the number of INSs in the
RASS/BSC should be held in some doubt.

The INSs which were first cataloged in the RASS/BSC all used ROSAT/HRI
localizations ($\sim$1\arcsec) to exclude nearby possible
counterparts, and so concentrated on the brightest sources in the
RASS.  Confusion of optical sources due to the large error-circles of
the RASS/BSC hampers INS identification as we discuss in the following
section.

In the present work, we obtained a selection of candidate INSs through
an expanded application of statistical cross identification
\cite{xid}, which we summarize in \S\ref{sec:xid} using optical, radio
and infrared catalogs.  We then present the list of thirty-two
candidate INSs, and proceed to exclude sources as INSs when they are
shown to be X-ray variable, or have \lxlbol$<$\ee{4}.  Some are
excluded from existing identifications of X-ray variable sources in
the literature.  \chandra\ X-ray observations which provide arcsec
localizations which result in counterpart identifications -- or X-ray
flux upper-limits, indicating a variable X-ray source -- are described
in \S\ref{sec:obs}; we also describe examination of archival X-ray
observations of four sources with ROSAT/HRI, which provide
localizations which are also sufficient for identification, or X-ray
upper limits for variable soruces.  In \S\ref{sec:estimate}, we
estimate the upper-limit on the number of INSs detected in the
RASS/BSC based on our analysis results, and compare this limit with a
naive and optimistic model for a cooling NS source population.  A
summary and conclusions are given in \S\ref{sec:con}.

\section{Source Confusion in Previous Work}
\label{sec:confusion}

The observational problem of discovering INSs (for a recent review,
see \citenp{treves00}) from the RASS/BSC is severely hampered by
source confusion
\cite{motch97c,zickgraf97,bade98,danner98a,danner98b,thomas98}.  For
example, the average number of optical sources in the USNO-A2 catalog
in a 3$\sigma$ uncertainty region of a RASS/BSC source is $\sim$3, and
is often much greater in the Galactic plane.  However, at such source
densities, an arcsec localization which places the X-ray source
coincident with an off-band counterpart results in a probability of
random alignment of $\sim\exp{(-\rho\pi (1'')^2)}=$0.3\%.  Thus, an
arcsec localization coincident with an optical source in the USNO-A2
argues for association, with a probability of spurious association at
the level of 0.3\%.  For the faintest X-ray sources in the RASS/BSC,
the implied $L_X/L_{\rm opt}$ for the faintest optical sources in
USNO-A2 is $\sim$few -- comparable to what is expected for AGN
(0.1-10).  At higher optical fluxes, the values of $L_X/L_{\rm opt}$
approach that of stellar coronae (\ee{-4}--\ee{-3}).

Previous INS searches took the following approach in identifying new
INSs: (1) examine off-band source catalogs, images, or spectroscopic
surveys for objects which are spatially close to the X-ray source
position; (2) all such off-band sources are individually evaluated to
be either a plausible counterpart or not; (3) if a plausible
counterpart is found, the X-ray source is considered identified as a
class which is not an INS.

This approach is a reasonable means for searching for INSs.  However,
none of the previous works quantitatively assess their confusion rate
-- specifically: if an INS were placed in one of their fields, what is
the probability that they would mis-identify it with a plausible, but
incorrect, counterpart?

While the discovered INSs place a lower-limit on the number of INSs in
the sky, the upper-limit on INSs in the sky cannot be quantified
without knowing the probability of mis-identification. We therefore
question the reliability of upper-limits derived from these works.
This statistical question is the basis upon which all conclusions
regarding the INS population rests
\cite{neuhauser99b,treves00,popov00}.

Neuh\"auser \etal (\citenp*{neuhauser99b}; N99 hereafter) examined the
observed log N-log S curve of INSs, to compare the relative
contribution to this curve of INSs powered by Bondi-Hoyle accretion
from the inter-stellar medium (ISM) vs. those powered by cooling from
their recent ($<$\ee{6} yr) SNe.  Citing previous work
\cite{bade98,motch97c,motch97a}, N99 estimated a 2\%
mis-identification rate; however, we find no quantitative support for
this estimation in any of the cited works.  The limits derived by N99
for the number of cooling or accreting INSs therefore function only as
lower-limits.

We therefore set out in the present work to understand our
mis-identification rate, so that we may robustly place an upper limit
on the number of INSs in the ROSAT Bright Source Catalog.

\section{Candidate Source Selection}
\label{sec:xid}

To produce this selection, we performed a statistical
cross-association between the RASS/BSC X-ray sources and sources from
three large-area catalogs:

\begin{itemize}
\item The National Radio Astronomy Observatory
\footnote{The National Radio Astronomy Observatory is a facility of
the National Science Foundation operated under cooperative agreement
by Associated Universities, Inc.} (NRAO) Very Large Array (VLA) Sky
Survey (NVSS) catalog \cite{condon98}, which contains 1.4~GHz radio
sources observed with the Very Large Array radio observatory between
1993 September and 1996 October, with a sensitivity of $\sim$0.3 mJy,
containing $\sim$2\tee{6} objects. 

\item The Infrared Astronomical Satellite (IRAS) Point Source Catalog
(v. 2.0) contains $\sim$250,000 objects.  We use the IR source
position, positional uncertainty, and the 12 $\mu$m flux density for
source flux. The position was taken as the source centroid (precessed
from B1950.0 to J2000.0 using the IDL routine jprecess from the ASTRO
package), and the positional uncertainty was taken as the major axis
of the elliptical positional uncertainties. 

\item The United States Naval Observatory A2.0 (\usno) catalog
contains $\approx$5\tee{8} optical sources, found by scanning optical
plates from the Palomar All Sky Survey (POSS-I) at $\delta>-30^\circ$,
and from the Science Research Council $J$ (SRC-J) and European
Southern Observatory (ESO-R) survey at declinations below that.  We
make use of the optical source position, positional uncertainty, and
quoted $B$ magnitude. The positional uncertainty of the \usno\ catalog
is generally $\sim$0.25\arcsec, although for objects brighter than 11
magnitude, which saturate the plates, the astrometry is accurate to
$\approx$2\arcsec. We adopted 1\arcsec\ as the positional uncertainty
for each optical source.  Based upon an initial investigation, we
considered only \usno\ sources which were $<$75\arcsec\ from the X-ray
source position.  The photometric accuracy is estimated as internally
0.15 mag, with systematic errors of 0.25-0.50 magnitudes); however,
for our purposes, this is of sufficient accuracy to be useful.  We
excluded objects with $B>20.0$ from consideration.
\end{itemize}

The approach is described in detail for a cross-association with a
single catalog (USNO-A2) elsewhere \cite{xid}.  Here, we summarize the
relevant descriptive elements for the present multi-catalog analysis
which produced the initial candidate INS source list.

As we are limited to declinations $\delta>-39\deg$ (the lower-limit of
NVSS) we have a total of 15,205 RASS/BSC sources. First, we collect
all sources in the three off-band catalogs which are within
150\arcsec\ (except for USNO-A2, where we use 75\arcsec) from the
RASS/BSC positions.  For each X-ray source $i$ and off-band source $j$
pair we calculate a figure of merit $LR_{i,j; C}$ using a function
which is particular to the off-band source catalog $C$:

\begin{equation}
LR_{i, j; C}=
\frac{\exp^{-r_{(i,j)}^2/2\sigma^2_{(i,j)}}}{\sigma_{(i,j)}\; N(>F_j; C)}
\end{equation}

\noindent where $r_{(i,j)}$ is the separation between the X-ray source
and the catalog object; $\sigma_{(i,j)}$ is the uncertainty in
$r_{(i,j)}$, found by summing the quadrature of the uncertainty in the
X-ray position and the off-band catalog object position (taking the
largest uncertainty in the case of NVSS, where the uncertainties are
typically elliptical); and $N(>F_j; C)$ is the fraction of sources in
catalog $C$ (actually, in the ``background fields'', see below) with
greater fluxes than observed from source $j$.

In addition to using the single-object catalogs alone, we combined
objects in groups of up to three (singles, doubles and triples), so
that new ``complex'' catalogs of multi-source objects were used. For
example: we formed complex catalogs of USNO-A2/USNO-A2, USNO-A2/IRAS,
USNO-A2/NVSS, NVSS/NVSS and so-on, producing a total of six
``doubles'' catalogs.  For each double-object, the value of $LR$ is
the product of the individual objects which make up the double-object;
if one USNO-A2 source had LR=5, and a second had LR=2, then the LR for
the ``complex'' object comprised of both sources had LR=20; and
similarly for triple-objects.  

We then use 24 off-source positions (the ``background fields''),
offset in a 5$\times$5 grid with separations of 300\arcsec\ and radii
of 150\arcsec (except USNO-A2, where we use 75\arcsec).  We exclude
background fields which contain RASS/BSC sources.  We calculate
$LR_{i, j; C}$ for those ``pairs'' using now the center of the blank
fields as the X-ray ``source'' position.  Then, we calculate a
``reliability'' R for each pair $i, j; C$:

\begin{equation}
\label{eq:R}
R_{i, j}(LR_{i,j;C}) =  \frac{N_{\rm src}(LR_{i, j; C}) - N_{\rm
background}(LR_{i, j; C})}{N_{\rm src}(LR_{i, j; C})}
\end{equation}

\noindent where $N_{\rm src}(LR_{i, j;C})$ is the number of objects
in the source fields which had a value $LR$ within a $\delta LR$ of
that obtained for the pair $i, j$, and $N_{\rm background}(LR_{i,
j;C})$ is the same, but in background fields.  The value $R$ is a pure
probability -- the probability that the candidate counterpart $j$ is
not a background object, but is associated with the X-ray source in
some way (and not necessarily as the X-ray emitter).  $R$ is
independent of the source catalog $C$ or the nature of the object $j$,
and so we now can mix the values of $R$ for the same X-ray source $i$
to produce the probability that off-band source $j$ is associated with
the X-ray source exclusive of other off-band sources in the field:

\begin{equation}
P_{i,j {\rm id}} =    \frac{R_{i, j} \Pi_{j'\neq j} (1 - R_{i, j'}) }{K}
\end{equation}

\noindent Similarly, the probability that none of the objects in the
field are associated with the X-ray source $i$ is: 

\begin{equation}
P_{i, {\rm no-id}} =    \frac{\Pi_{j} (1 - R_{i, j}) }{K}
\end{equation}

\noindent and the normalization $K$ is: 

\begin{equation}
K = \Pi_{j} (1 - R_{i, j}) + \Sigma_j \, R_{i, j} \Pi_{j'\neq j} (1 - R_{i, j'})
\end{equation}

When $P_{i, no-id}$ is close to 1, none of the off-band objects in the
field have properties which demonstrate a statistical excess in X-ray
fields over background fields -- that is, the off-band objects in the
field are beyond the confusion limit.  As found previously \cite{xid},
typical uncertainties in the values of \pid\ and \pnoid\ are
$\sim$2\%.

In performing this analysis, we also inserted 150 ``control'' sources
among the source fields, set randomly about the sky in proportion to
the local X-ray source density within our survey area.  These act as a
control experiment -- since they are not real sources, they will have
no detectable off-band counterparts in the all-sky catalogs, just as
for INSs. This permits us to evaluate the efficiency with which our
statistical selection will identify real INSs. Of the 150 ``control''
sources, 29 were found with \pnoid$>$0.90 by this procedure; the
remaining 80\% of the control sources had \pnoid$<$0.90, due to
confusion with nearby optical sources.

In Fig.~\ref{fig:control}a, we show the spatial distribution of the
control fields in Galactic coordinates. In Fig.~\ref{fig:control}b, we
show the 29 fields found with \pnoid$>$0.90.  While the control fields
are evenly distributed about the sky in our survey region, the
``found'' fields clearly avoid the Galactic plane.  This is due to the
fact that so many optical point sources populate the plane, that the
fields are confused with spurious (low significance) sources
(Fig.~\ref{fig:control}c).  

In Fig.~\ref{fig:control}d, we show the cumulative distribution
$N(<|b|)$ of the control sources with \pnoid$>$0.90, in comparison
with one expected from an equal area distribution in $|b|$ (taking
into account the declination constraint for our survey area, which
makes a difference only at the $\sim$few percent level in the
cumulative distribution).  Note that no control fields are found with
$|b|<20 \deg$.  Using a K-S test \cite{press}, the detected control
sources are inconsistent with an equal area distribution (KS
probability of prob$_{\rm KS}$=3\tee{-6}), which we attribute to
source confusion in the Galactic plane.  Modeling the absent sources
as a correction of $F(b)=sin^n(b)$, we find 90\% confidence limits on
$n$ of 1.25-3.8, with a best value of $n=2.2$.

\subsection{Source Selection}

We began with objects for which $P_{\rm no-id}>0.90$ (60 objects).
From these, we excluded all objects in which the RASS/BSC hardness
ratio HR1 indicated an effective temperature $>$200 eV (28 sources),
which are too spectrally hard to be cooling or Bondi-accreting INSs.
In the RASS/BSC, HR1 is defined in the usual fashion (=(B-A)/(B+A)),
where B is the number of counts detected in the 0.5-2.0 keV band, and
A is the number of counts detected in the 0.1-0.4 keV band), and we
require HR$>$0.  Because Bondi-accreting or cooling INSs are expected
to have lower effective temperatures than 200 eV, we expect to lose no
INSs from this cut (previously known INSs also have effective
temperatures $<$200 eV).  The thirty-two candidate sources are listed
in Table~\ref{tab:inscandidates}, along with their number in the ROSAT
Bright Source (RBS) catalog \cite{rbs1,rbs2,rbs3}, and other
particulars as we describe in this section.

We show in Table~\ref{tab:ins} the calculated values of \pid\ for the
seven known INSs listed in \acite{treves00}, plus one more identified
more recently \cite{zampieri01}.  Three are at declination below the
NVSS limit, and so were not included in our analysis.  Three have
optical sources in the field which have probabilities of being the
counterparts which would have selected them out of our sample.  Two
more, as we stated above, have a \pnoid$\sim$1, such that they would
have been included in our sample for \chandra\ observations.  This
lends confidence that our approach is capable of selecting INSs,
although -- as expected -- source confusion hampers discovery.

\subsection{Source Classification}
\label{sec:class}

We classify the thirty-two sources into one of three classes: (1) an
INS; (2) not an INS; (3) undetermined.  These classifications are
listed in column 6 of Table~\ref{tab:inscandidates}, with the relevant
reference for their classification in column 7.  Here, we discuss the
classifications each in turn.

Two of these sources are previously identified INSs. 

A source is definitively not an INS when either: (a) it is localized
to $\sim$1\arcsec\ (with either a \chandra\ or ROSAT/HRI
localization), and there is a spatially coincident optical counterpart
with \lxlbol$<$\ee{4}; or (b) the X-ray source is variable on any
timescale.

Finally we regard the classification as undetermined when the X-ray
localization is not at the 1\arcsec\ precision level, and when the
X-ray source has not been shown to vary in intensity.  

We first examined the SIMBAD-listed identifications and
classifications for these sources (see Table~\ref{tab:inscandidates}).
From these sources, we examined the literature for observations in
which the X-ray source was shown to vary in intensity.  We found five
such sources in our list, which we classify as not INSs.

We obtained 1 ksec \chandra/HRC-S observations of eight selected
fields.  As we describe in \S~\ref{sec:chandra}, in all cases we found
either off-band counterparts at the arcsec \chandra\ positions, or
that the X-ray sources were variable, resulting in a classification of
``not an INSs''.

Four of the remaining sources had ROSAT/HRI observations in the
archive.  Using these, we show that either the sources were variable,
or could be identified with an off-band counterpart using the
$\sim$few arcsec localization (see \S~\ref{sec:rosat}) and are
therefore not INSs.

We found thirteen sources with either previous identifications but no
evidence of X-ray variability; or for which there were nearby
(\approxlt 30\arcsec) off-band sources which belong to classes which
are known X-ray emitters, which we list as possible counterparts.  We
classify these sources as ``undetermined''.

Following these classifications, we find: 2 INSs, 17 sources which are
not INSs, and 13 with undetermined classification.

\section{X-ray Observations and Analyses}
\label{sec:xrayobs}

In this section, we describe the X-ray observations on which we rely
to provide the arcsec localizations, demonstration of X-ray
variability, and subsequent classifications as ``not an INS''.  The
\chandra\ observations are part of an observing program to search for
INSs, and so we describe these observations in some detail in
\S~\ref{sec:chandra}.  The \rosat\ observations, taken from the public
archive, are more summarily described in \S~\ref{sec:rosat}.

\subsection{\chandra\ Observations and Analysis}
\label{sec:obs} \label{sec:chandra}

All observations were performed with the HRC-S in imaging mode.
Details of each observation are in Table~\ref{tab:ins}.  All analyses
were performed with CIAO v2.2 \footnote{http://asc.harvard.edu/ciao/}.
\chandra\ source localizations were performed with {\tt wavdetect}, by
binning the data by a factor of 8 (1.05\arcsec\ per bin), and
searching for sources on scales of 1, 2, 4, 8, and 16 bins
(1.05-16.9\arcsec), with significance threshold of \ee{-6}.
Upper-limits to the average countrate, when no source is detected, are
found by rebinning the area within 2\arcmin\ of the \rosat\ position
to 2.1\arcsec\ wide bins, and finding the greatest number of
counts/bin.  All detected X-ray sources were found on the smallest
wavelet scale (1.05\arcsec) indicating that none were extended.

Here, we detail the results of individual X-ray source searches, and
individual optical counterpart identifications.  Where we find an
X-ray source in our \chandra\ data, we search for optical counterparts
in the Digitized Sky Survey (DSS), IR counterparts in the 2MASS 1st and
2nd incremental databases (which cover $\sim$40\% of the sky, for
which astrometry and photometry are available) the 2MASS quicklook
image database (which cover 100\% of the sky, but without astrometry
and photometry), and in the Digitized Palomar Sky Survey (DPOSS;
\citenp{dposs}). 

As the RASS/BSC error circles are typically $>$8\arcsec (1$\sigma$)
and there is typically 1 optical source in these fields localized to
1\arcsec, the {\em a priori} probability that any X-ray source will
spatially coincide with an unassociated optical source is $<$0.5\%. We
therefore do not expect any spurious spatial associations with optical
sources.

We estimate fluxes assuming a steep photon spectral slope of
$\alpha=3$ and X-ray column \nh=0, for which 1 \chandra/HRC-S count =
1.4\tee{-12} erg \perval{cm}{2} (0.5-2.0 keV).

The analysis results for the individual targets are as follows. 

{\bf 1RXS~J020317.5$-$243832}.  Two X-ray sources are found in this
field, both of which lie close to optical point sources in the DSS.
As this offers the opportunity to do much better astrometry, we
handled the absolute astrometry of this source slightly differently
from our other targets.  We correct the X-ray aspect according to the
Chandra X-ray Science Center prescription
\footnote{http://asc.harvard.edu/calASPECT/fix\_offset/fix\_offset.cgi},
which has typical 1$\sigma$ systematic uncertainty of 0.6\arcsec.  We
rebinned the X-ray data by only a factor of 2, and used {\tt
wavedetect} to obtain localizations, which were relatively accurate
with a precision of 0.04\arcsec\ for the and 0.09\arcsec for
1RXS~J0203$-$2438 and the second object, respectively.  Both can be
identified with optical sources in the DSS, as well as in a 5-min
integration image taken with ESI at Keck II shown in
Fig.~\ref{fig:0203esiimg}.  To perform absolute astrometry, a 30-sec
exposure of the field was also obtained with ESI/Keck II; six Guide
Star Catalog II stars over this field were used to provide an absolute
astrometric registration of 0.25\arcsec (1\sig); the 5-min exposure
was then registered relative to the 30-sec exposure.  Finally, the
second X-ray source in the field was assumed to be exactly coincident
with an optical point source in the field -- which coincided with the
X-ray position to within \chandra\ astrometric errors (0.6\arcsec).
The \chandra\ source corresponding to the RASS/BSC object is
CXO~020317.626-243837.8, with an absolute uncertainty in its position
of \ppm0.3\arcsec.  The corresponding optical source is near the plate
limit on the DSS blue plate, and is not listed in the USNO-A2 catalog,
nor is it detected in the 2MASS quicklook images.  Taking the DSS
plate limit to be $B$=20.8 (the faintest of 10 objects within
100\arcsec\ of the X-ray source, listed in USNO-A2), and assuming the
optical source to have a magnitude equal to this limit, this sets an
$L_X/L_B\approx2$, with fractional uncertainties at the level of 50\%,
due to uncertainty in the optical flux near the plate limit.  This is
consistent with an AGN origin for this object, which we offer as a
tentative classification.

We observed the optical counterpart with the 10~m Keck II telescope
and Echelle Spectrograph and Imager (ESI) on 5~November 2002 UT.  In
its echelle mode ESI provides uniform 11.5 km~s$^{-1}$~pixel$^{-1}$
spectral resolution in the 4000--11,000 angstrom wavelength range.  A
single 600~s spectrum of the source reveals a broadband continuum with
no prominent emission lines, and thus does not help much to clarify
either the nature of the source or its distance or redshift.


{\bf 1RXSJ~024528.9+262039}. The \chandra\ X-ray source is spatially
distant (22.7\arcsec, 2.8$\sigma$) from the RASS/BSC position, and is
coincident with a bright IR source, detected in 2MASS and listed in
USNO-A2.  Our statistical analysis had placed a probability of unique
identification with the USNO-A2 object of $P_{\rm id}$=0.088, due
large distance from the ROSAT X-ray position.  The corresponding USNO
object has $B=14.8$, so $J-K$=0.93 and $B-K$=6.2.  The colors are
appropriate to a late M-type star (M3 has $B-K$=6.1, $J-K$=1.07; M4
has $B-K$=6.43, $J-K$=0.90, well within errors; \citenp{zombeck}).
For the $V-J$ and bolometric correction of an M3 star (B.C. = $-2.03$;
$V-J$=3.66), we find \lxlbol=6\tee{-3}.

We tentatively classify this X-ray source as a coronally active star of
spectral type M2-M4. The absolute J magnitude is then $M_J=6.98-8.34$
\cite{hawley02}.  The implied distance modulus is
$m_J-M_J$=1.11-2.47, or a distance between 16 and 31 pc.  The implied
X-ray luminosity is (2-6)\tee{27} \cgslum.

{\bf 1RXS~J115309.7+545636}.  There is no X-ray source detected in the
\chandra\ field.  We find an X-ray average countrate upper-limit of
$<$4.7 c/ksec, and $F_x<$7\tee{-15} \cgsflux.  At the time of
analysis, there was an undocumented hot pixel in the detector, which
was $>$2\arcmin\ from the RASS/BSC source position, and which does not
affect our analysis.

The X-ray source was present in pointed observations with ROSAT/PSPC
on 1996 June 25, listed in the WGA catalog \cite{white94} with a
countrate of 27 c/ksec, approximately a factor 5 fainter than during
the RASS (130\ppm20 c/ksec).  Thus, the object is variable on at least
a 5-year timescale.  At the fainter flux of the pointed ROSAT
observation, this object would have been only marginally detected
(with $\sim$6 counts) with the HRC-S.

{\bf 1RXS~J122940.6+181645}.  There is no X-ray source detected in the
\chandra\ field.  Based on the source hardness ratio and PSPC
countrate, this was predicted to be the brightest object in the
survey, with 400 counts/ksec.  We find no 2\arcsec\ regions within a 2
\arcmin\ radius of the RASS position with more than 3 counts,
corresponding to an observed average X-ray countrate upper-limit of
$\leq$2.7 c/ksec, and $F_X\leq$4\tee{-15} \cgsflux.  The observed
X-ray limit indicates this source has faded by 2 orders of magnitude
from the average X-ray flux observed during the RASS.

The ROSAT source has been tentatively identified as a BL Lac-type
object RBS~1116 of unknown redshift \cite{schwope00}, with $V$=20.5
found from folding a low resolution spectrum with the sensitivity
curve of a $V$ filter, in the process of the Hamburg survey for
quasars in the RASS/BSC \cite{bade98}.  In a 3\arcsec\ circle centered
at the location of this optical source given in \acite{schwope00}, we
find 1 count (with 1.0 background counts expected); this implies an
upper-limit to the average countrate of the source of $<$ 5
counts/ksec or $F_X<$7\tee{-15} (96.6\% confidence), and
$F_X/F_V<0.3$.

{\bf 1RXS~J132833.1-365425}. This was the brightest detected X-ray
source in this survey, with 687\ppm25 c/ksec, a factor of 10 brighter
than predicted from the RASS/BSC countrate.  Visual examination of the
X-ray lightcurve showed no evidence for variability on a 1 or 100
second timescale.  The X-ray source position (Fig.~\ref{fig:1328}) is
coincident with an IR source in the quicklook 2MASS images (Epoch
1999.45; 2MASS catalog photometry and astrometry is not yet
available); we associate the \chandra\ X-ray source with this IR
source.  The nearest DSS object is $\sim$4.3\arcsec\ \ppm1\arcsec\
away, in an image taken in epoch 1975.27; although the source is well
above the DSS plate limit, it is not listed in USNO-A2, for reasons
which are unclear. The source is listed in the GSC 2.2, with
$B$=15.42\ppm0.17. A plate taken during the Second Epoch UK Schmidt
survey shows that the earlier epoch source has measurable proper
motion, consistent with that object being at the location of the 2MASS
source at the Epoch of the 2MASS observation. The implied proper
motion is 165 milliarcsec/yr, with a 1$\sigma$ 25\% uncertainty due to
\chandra\ absolute astrometric uncertainties, indicating the source is
a nearby, low-mass star.

{\bf 1RXS~J145010.6+655944}.  There is a faint source visible in the
DSS2 blue plate, within 1\arcsec\ of the \chandra\ X-ray source
position.  However, it is not detected on the red plate, and it is not
included in USNO-A2.  It is also detected by DPOSS
(Fig.~\ref{fig:1450}).  There is no 2MASS source at the X-ray position
in the 2MASS quicklook images. An optical spectrum of this source was
taken with Keck in May 2002 (Fig.~\ref{fig:lris}).  The spectrum shows
several narrow emission lines.  We identify the object as a
cataclysmic variable (CV) on the basis of its H$\alpha$ line (compare,
for example, with spectrum from SDSS 1555, \citenp{szkody02}).

{\bf 1RXS~J145234.9+323536}.  No X-ray point source was detected in
the \chandra\ observation.  This position on the sky has not otherwise
been observed with the ROSAT, ASCA, or SAX X-ray observatories.
Binning the 2\arcmin\ circle around the source position into
2.1\arcsec\ pixels, the largest number of counts per pixel was 3, for
an upper-limit to the average countrate of $<$2.3 c/ksec, or
$F_X<$3\tee{-15} \cgsflux. This is a factor of 44 below the predicted
countrate.

{\bf 1RXS~J163910.7+565637}. The X-ray source is associated spatially
with an optical source (Fig.~\ref{fig:1639}) near the center of the
error box in DPOSS. The source is also observed near the survey limit
in 2MASS quicklook images. The source is detected in the DSS blue and
red plates, but it is not listed in USNO-A2, possibly due to
extendedness. It is, however, listed in the HST Guide Star Catalog 2.2
as source N11231206454, with $B$=17.8\ppm0.42; this corresponds to a
flux ratio $F_X/F_B\sim0.1$, comparable to those expected for AGN. An
optical spectrum taken with Keck/LRIS in May 2002 showed a broad
emission line ($\sim$4000 km \perval{s}{-1}) \ref{fig:lris} at
7412\AA.  If this is identified with MgII, then the redshift of the
source is $z=1.65$\ppm0.01. Assuming a Hubble constant $H_0$=65
km~\perval{s}{-1}, $\Omega_\Lambda=0.7$ flat universe, the intrinsic
X-ray luminosity is 1.4\tee{45} \cgslum (in the observed 0.5-2.0 keV
band).  Assuming a K-correction of 2~mag, the absolute $B$ magnitude
is $M_B=-29.8$.

\subsubsection{Comparison of \chandra\ results, RASS/BSC Positions,
and USNO-A2 Sources in the Digital Sky Survey}

In Fig.~\ref{fig:dss}, we show the eight fields of our INS candidates,
taken from the Digital Sky Survey (DSS).  The RASS/BSC error circle
(3$\sigma$) is shown for each source, as well as the \chandra\
localization for the five sources which were detected.  Of the three
X-ray sources associated with AGN, two (1RXS~J0203-2438,
1RXS~J1450+6559) have optical counterparts which were at or below the
DSS plate limit; the third (1RXS~J1639+5656) has an optical
counterpart which is well above the plate limit but which was not
listed in USNO-A2, possibly due to its extendedness.  Finally, of the
two sources which are associated with stars, one (1RXS~J0245+2620) was
listed in USNO-A2, but was close to 3$\sigma$ from the RASS/BSC
position, and so had a low probability of association (\pid=0.088).
The other (1RXS~J1328-3654), which was $<1\sigma$ from the RASS/BSC
position, is not listed in USNO-A2.

Had the optical counterpart of 1RXS~J1639+5656 been included in
USNO-A2, we find that it would have had \pid$\sim$0.70, and would not
have been included in our selection.  Similarly, the optical
counterpart to 1RXS~J1328-3654, had it been included in USNO-A2, would
have had \pid=0.85.  As such, neither would have ended up in our
selection as INS candidates.  The other optical counterparts which did
not appear in USNO-A2 were well below the optical confusion limit, and
would have had \pid=0.

\subsection{ROSAT HRI Observations and Analysis} 
\label{sec:rosat}
Four sources in our list have archived ROSAT/HRI observations.  We
examined each in turn, using data from the HEASARC ROSAT archive at
GSFC.  We took positions and countrates from the first ROSAT Source
Catalog of Pointed Observations with the High Resolution
Imager\footnote{ROSAT Consortium, ROSAT News No. 74, 2001 Aug 9},
while estimating upper-limits to countrates from the raw data.
Although the X-ray focusing capability of the ROSAT/HRI permits
relative X-ray localization to 1\arcsec, there is a systematic
uncertinaty in the absolute positions derived from ROSAT/HRI
observations at the level of $\sim3$\arcsec\ due to a boresight
uncertainty, which must be kept in mind during analysis.

{\bf 1RXS J094432.8+573544.} A 3.4~ksec HRI observation (RH704086N00)
detects both this and a second X-ray source, with $\sim$arcsec
precision (at 09h44m31.s8+57d35m38s and 09h44m01.s9+57d3m210s
respectively).  Comparing these sources with DPOSS images finds that
they can be re-registered with an offset consistent with the known
ROSAT/HRI boresight uncertainty.  Source 2 corresponds to
USNOA2~094402.52+573209.2 ($B=17.4$, $V=17.7$); re-registering the
X-ray source to 09h44m32.s42+57d35m37.s2 finds that it corresponds
with DPOSS~094432.42+573534.9 with (g,r,i)=19.68,20.59,20.11), with an
offset of 2.9\arcsec.  There are four sources in the DPOSS catalog
with $g>20.0$m, within a 120\arcsec$\times$120\arcsec box of this
source.  The probability of chance alignment at 2.9\arcsec\ of one of
these sources is 0.7\%.  The optical counterpart is too bright to
correspond to an INS, and so we identify the X-ray source as not an
INS.  Its \lxlopt\ ratio is comparable to that of known AGN, and
therefore this source may be an AGN.

{\bf 1RXS J130547.2+641252} During the one 1.6 ksec HRI observation
(RH704083N00), the source was undetected ($<$3 counts) within a
60\arcsec\ radius of the RASS/BSC position.  The RASS/BSC countrate
(0.17\ppm0.02 PSPC c/s) corresponds to 0.054 HRI c/s (assuming a
power-law photon spectrum $\alpha=3$), or 87 counts during this
observation, indicating that the source has faded by a factor of
\approxgt30.  Based on this variability, we  identify the
source as not an INS.

{\bf 1RXS J130753.6+535137} This X-ray source has been identified
(from spatial coincidence) with the very well studied CV V* EV UMa.  A
11.9ksec HRI observation (RH300382N00) does not detect the X-ray
source within 60\arcsec\ of the RASS/BSC position ($<$ 5 counts).
Based on the RASS/BSC countrate (1.86\ppm0.06 PSPC c/s), the predicted
number of counts in the HRI observation is 7100, indicating that the
source has faded by a factor \approxgt1400.  Based on this
variability, we identify the source as not an INS.

{\bf 1RXS J163421.2+570933} We examined HRI observations centered on
this source (RH201944N00 and RH202224N00, epochs 1995.27 and 1996.24
respectively) for which the source is focused within the central
1\arcmin\ of the observation.  The X-ray source is detected in both
observations.  In the latter observation, its arcsec localization was
16h34m21.17s+57d09m38s.  We examined the 2MASS J-band quick-look
image (epoch 1999.36) and DSS V-band images from two epochs (epochs
1955.32, 1991.50).  We find a high proper-motion binary, with two
components separated by 37\arcsec.  The southern component
(2MASS~J163420.44+570944.0, -- a (J,H,K)=8.50, 8.04, 7.77m, with
colors corresponding to a K5-7 dwarf, or a K2 giant) -- moves with
($\mu_\alpha,\mu_\delta$)=($-$1135\ppm5, +1170\ppm30) mas/yr, which
places it at 16h34m20.87s+57d09m40.3s at the epoch of the ROSAT/HRI
observation, which is 3.4\arcsec\ away from the ROSAT/HRI position,
within the boresight correction uncertainties.  The northern component
is 2MASS~163421.6+57108.2, with (J, H, K)=(14.09, 14.08, 14.07) -- a
flat spectrum, possibly corresponding to an A-type star.  There are 15
$K<7.77$ point sources in 2MASS, within 1 degree of this position, and
the probability of one falling at $<$3.4\arcsec\ from the ROSAT
position randomly is 1.3\tee{-5}.  Based on the low likelihood of
positional coincidence, we associate the X-ray source with
2MASS~J163420.44+570944.0 and identify it as not an INS. 

\section{Estimation of the Number of INSs Detected in the RASS/BSC}

\label{sec:estimate}

To place a limit on the total number of INSs detected among the 18,811
X-ray sources in the RASS/BSC, we model the selection of our candidate
list, and correct for each step.

{\bf Step 1: Statistical selection of INSs}: As described in
\S~\ref{sec:xid}, we performed a statistical cross identification with
the intention of finding INSs from the absence of any likely off-band
counterparts, as quantified by \pnoid$>$0.90.  From the original
15,205 RASS/BSC sources above our declination cut, 60 X-ray sources
passed this selection.  Of $A=150$ ``control'' sources, $B=29$ passed
our \pnoid\ selection.  A real INS in our survey region has a
probability $p$ of of passing our INS selection; we represent this
value $p$ as a binomial distribution (Fig.~\ref{fig:pxid}):
\begin{equation}
\label{eq:bin}
P_{\rm XID}(p) = \frac{ p^{B}(1-p)^{A-B}} {\int_0^1 p^{B}(1-p)^{A-B}\;
dp}
\end{equation}

{\bf Step 2, Spectral Hardness Cut}: We selected a total $T=32$ (from
60) X-ray sources with spectral hardnesses corresponding to effective
temperatures \kteff$<$200~eV, which we assume loses no INSs since
cooling and accreting (non-magnetic) INSs have effective temperatures
below this value \cite{popov00b}.

{\bf Step 3: Candidate Classification}: After examining the
literature, X-ray variability history, and new arc-sec localizations,
we classified all $T=32$ candidates as either an ``INS'' (2 sources),
``not an INS'' (17) or ``undetermined'' ($BG=$13).  We discovered no
new INSs among our candidates.  However, any (or all) of the $BG$
objects may well be INSs.  We require a correction, which is the
(un-normalized) probability $P_{BG}(N)$ that a number $N$ sources in
the $T=32$ selection are INSs, with only $N_{\rm INS, min}=2$ INSs so
identified:

\def\combfactor{
{\frac{   \left(\begin{array}{c} N \\ N_{\rm INS, min}  \end{array}\right)
\left( \begin{array}{c} T - N \\ T-BG-N_{\rm INS, min}  \end{array} \right) }
	{ \left( \begin{array}{c} T \\ T-BG  \end{array} \right) }}
}

\begin{equation}
\label{eq:comb}
P_{BG}(N)= {\combfactor}
\end{equation}

\noindent Here $N_{\rm INS, min}\leq N \leq N_{\rm INS, max} $ -- that
is, $N$ ranges from the minimum number of INSs known to be present in
our sample of $T$ objects to the maximum number of INSs which could be
present in the sample of $T$ objects ($N_{\rm INS, max}=N_{\rm INS,
min}+BG=19$). See Fig.~\ref{fig:pbg}. The function $
\left(\begin{array}{c} x \\ y \end{array} \right) $ is the familiar
combinatorial factor $x!/(y!(x-y)!)$.

These selections combined produce the unnormalized probability $P_{\rm
INS}(M')$ that the total number of INSs in our survey field is $M'$: 


\begin{equation}
P_{\rm INS, un-normalized}(M')    = 
\sum_{N=N_{\rm INS, min}}^{{\rm min}(M', N_{\rm INS, max})}    \:
\left( \frac{P_{\rm BG}(N)} {\sum_{N=N_{\rm INS, min}}^{{\rm min}(M', N_{\rm INS, max})}      \:  P_{\rm BG}(N) }    \,    \int_0^1 P_{\rm XID}(p)\,  
\frac{p^N (1-p)^{M'-N} }{\int_0^1 p^N (1-p)^{M'-N}\; dp}\; dp\right)  
\end{equation}

\noindent where ${\rm min}(A,B)$ takes the lesser value of $A$ and
$B$.  To produce the probability that there are $\geq M$ INSs in our
survey field, we sum: 

\begin{equation}
\label{eq:pins}
P_{\rm INS}(\geq M)   = \sum^\infty_{M'=M} \frac{\: P_{\rm INS, un-normalized}(M')}
{\left(\sum^\infty_{M'=N_{\rm INS, min}} \:P_{\rm INS, un-normalized}(M') \right)}
\end{equation}

We begin the summation for the number of INSs in our fields at $M=2$
since this is the number of INSs we would have found with arc~sec
\chandra\ localizations.  This results in upper-limits on the number
of INSs detected in the RASS/BSC in our survey area of 56 (90\%
confidence -- that is, $P_{\rm INS}(\geq 56)=0.1$) and 87 (99\%
confidence).  These limits are consistent with the number of INSs
previously known to be in our survey area (five).

A perhaps more intuitive (but less precise) way to obtain this limit
is the following.  We ultimately found 2 INSs.  The combinatorial
probability (Eq.~\ref{eq:comb}, Fig.~\ref{fig:pbg}) shows that there
can be at most 9 INSs (90\% confidence) in the selection of 32.  Our
control fields showed (Eq.~\ref{eq:bin}, Fig.~\ref{fig:pxid}) that for
every 1 INS we find in our selection there are five INSs in our survey
area. Thus, the 90\% upper-limit of 9 in our selection becomes an
upper-limit of $5\times9=$45 in our survey area.  After adding an
additional 20\% due to the width of our ``control'' field selection
distribution (Fig.~{fig:pxid}), the 90\% upper-limit is 55 INSs in our
survey area, comparable to the upper-limit of 56 we found in the more
precise calculation.

Thirteen of the RASS/BSC sources in our sample did not have either an
$\sim$arcsec position, or demonstrate X-ray variability, to rise to
the level of being definitively classified as ``not an INS''.  All
thirteen had previously been examined by other workers, who had found
some nearby off-band source in the error-circle with which the X-ray
source may plausibly be identified.  While we do not accept these
associations due to the absence of quantitative association argument,
if one accepts these suggested associations, then the upper-limits on
the number of INSs in our survey field decrease to $<$34 and $<$56
(90\% and 99\% confidence, respectively).

\subsection{Full-Sky Number of INSs in the RASS/BSC, Assuming Isotropy}

Since the effects of the galactic-latitude dependent confusion are
already accounted for in our estimation of $p$, we do not need to
account for this in calculating the correction for the full-sky,
isotropic number of INSs in the RASS/BSC.  Correcting only for the
ratio of area coverage (a factor of 1.2) we find there are $<$67 and
$<$104 INSs at 90\% and 99\% confidence.  These limits are consistent
with the number of INSs in the full sky RASS/BSC (seven).

\subsection{Comparison of Observed Limit on INSs with a Naive and Optimistic Cooling NS Model}

Here, we compare this limit on the number of INSs in the RASS/BSC
with a naive and optimistic model model for cooling INSs in the
galaxy.  

We assume that such INSs are produced in the disk at a rate of
$\gamma_{-2}$ per 100 years, out to a radius of $R_{\rm disk}$=15 kpc;
that such INSs have a velocity perpendicular to the disk, which
remains constant in time, of a magnitude $V_{\rm perp.}$; that the
disk can be treated in this limit as an infinite plane; that the X-ray
luminosity is 2\tee{32} \cgslum\ in the ROSAT/PSPC passband (0.1-2.4
keV) for $\tau=$\ee{6}~yr, after which the luminosity is zero (see,
for example, \citenp{yakovlev01a}); and that the flux limit is
2\tee{-13} \cgsflux, which permits the detection of such INSs to a
distance of $\chi_{\rm lim}$=10.3~kpc.  In this model, the number of
detected NSs is:

\begin{equation}
N = \frac{\gamma_{-2} (0.01\,  {\rm yr}^{-1})}{\pi R_{\rm disk}^2 V_{\rm
perp.}}\; 2\pi \int_0^{min(V_{\rm perp.}\tau, \chi_{\rm lim})}
\int_0^{\sqrt{\chi_{\rm lim}^2-z^2}} R \, dR \, dz
\end{equation}

\noindent where $R$ is the distance of the INS from the observer in
the Galactic plane and $z$ is the distance of the INS above the
Galactic plane.  This model is optimistic, in that it assumes that the
INSs all have temperatures which are at the very upper-limit of
theoretical estimates.

There are several reasons why this model is naive.  First, it neglects
the effect of absorption in the plane of the galaxy, which cannot be
neglected for spectrally soft X-ray sources which would be observable
out to 10 kpc. Second, it parameterizes the velocity perpendicular to
the disk as a delta-function, while it is known that the radio pulsars
are observed to exhibit a range of velocities, and the NSs will
naturally travel with a distribution of angles relative to the disk.

Nonetheless, a limit of $<$67 INSs in the RASS/BSC in this model
produces a limit on $\gamma_{-2}<$0.025 (90\% confidence), for $V_{\rm
Perp.}<10$~kpc~\perval{Myr}{-1}, or a hot NS birthrate of 1 per 4000
years.  This is low compared with the estimated SNe rate
($\gamma_{-2}\sim$1), and most likely implies that the naive and
optimistic assumptions for this model are incorrect.
\acite{popov00b}, for example, have examined a cooling NS population
model which includes galactic absorption, showing its effect to be
important.

If we assume an $\alpha=3$ spectrum with a column density
\nh=3\tee{21}~\perval{cm}{-2} (slightly above the median all-sky
value; \citenp{dickey90}), the unabsorbed flux limit is substantially
greater: 9\tee{-12} \cgsflux, for which INSs could be detected out to
a distance of 1.5 kpc. Assuming an average spatial H density of
1~\perval{cm}{-3}, the column density out to such distances is
4.5\tee{21} -- comparable to our assumed \nh.  Under this assumption,
our limit on $\gamma_{-2}<1.1.$

Comparison of the present results with this naive and optimistic model
demonstrates that the limit on the number of INSs obtained is
sufficient for detailed comparison with the INS model population under
a range of assumptions (birth-rate, velocity distribution, NS cooling
models, galactic distribution).  We leave detailed comparison of these
observational results with realistic population models for future
work.

\section{Summary and Conclusions}
\label{sec:con}

We performed a selection of INS candidates, and examined them for
identification as INSs. We showed that this selection should identify
20\% of the INSs in the Galaxy; confusion of USNO-A2 sources in the
plane of the Galaxy is the major barrier efficient INSs
identification.  The selection found 32 candidate sources; of these,
two were previously known to be INSs, seventeen we determine to be not
INSs, and thirteen we leave as undetermined classification.  No new
INSs were discovered.

This results in limits for the number of INSs detected in the
ROSAT/BSC of $<$67 INSs full sky (90\% confidence) or $<$104 INSs full
sky (99\% confidence).  

Of eight RASS/BSC X-ray sources observed with \chandra, three have
faded, by factors of 5, 44, and 100; two are associated with
tentatively classified low-mass stars, one is a CV and the remaining
two are tentatively classified as AGN.  Of the four observed with the
ROSAT/HRI, we find that one faded by a factor of \approxgt30; one --
previously identified as a CV -- also faded by a factor of
\approxgt1400; the remaining two are a high-proper-motion binary, and
a source which may be an AGN.

The classification of four fading sources is uncertain.  One
(1RXS~J1229+1816) has previously been suggested to correspond to a
BL~Lac object in the field; we have no additional evidence of that,
except that BL Lac objects are known to vary dramatically in the
X-ray, and this X-ray source faded by a factor $>100$.

If most INSs are powered by NS cooling as in the first \ee{6}~yr
following supernova, then the assumption of an isotropic distribution
is well justified.  On the other hand, if most INSs are powered by
accretion from the ISM, one might expect a greater concentration of
INSs toward the plane, as the ISM has a scale-height of $\sim$300~pc;
however, because the mass accretion rate in Bondi accretion is
strongly velocity dependent $\dot{M}\propto v^{-3}$, and the mass
accretion rate in magnetically dominated accretion is dependent on the
velocity with the {\em opposite} sense as Bondi accretion
($\dot{M}\propto v^{1/3}$), as well as on the NS magnetic field
strength \cite{rutledgemagac,toropina01}, correcting for a
non-isotropic population is model-dependent, and we leave this for
future work.

Note that our statistical cross-association technique excludes sources
from consideration when the X-ray sources are associated with off-band
counterparts.  This is a reasonable approach if the INSs are not in
any way associated with -- for example -- optical sources.  If,
however, most INSs were born with low velocities, then they may stay
near their birthplace which may therefore be associated with bright
optical sources such as an OB association.  If this were the case,
then we would miss such INSs, and our limit does not apply to these.
If, however, NS formation or evolution processes give rise to a
peculiar velocity of $\sim$100 km\perval{s}{-1}, the INS will have
crossed a galactic disk width (100 pc) on a timescale comparable to
the cooling time (\ee{6} yr), so that it would not be associated with
its birthplace.

The present limits on the number of INSs in the RASS/BSC are
quantitatively comparable to those obtained from previous surveys
\cite[and references therein]{neuhauser99b}.  However, the limits from
previous surveys were found by estimating a confusion rate with a
spurious background AGN or stars; these estimates (typically $\sim$few
percent) were not based on any quantified analysis of the non-X-ray
AGN or stellar presence in background fields.  In the present
analysis, we quantified the confusion rate for INSs in our survey
field by using control sources, finding that $\sim$80\% of INSs would
be confused with background sources.  We therefore believe our
estimation of the confusion rate to be highly robust in comparison to
previous results, and our limits on the INSs in the RASS/BSC to be
robust as well.

The fact that all X-ray sources detected with \chandra\ or \rosat/HRI
yielded an optical/IR counterpart is enlightening.  It implies that
the faintest persistent X-ray sources of the RASS/BSC can be
successfully identified with 1\arcsec\ X-ray astrometry using USNO-A2
and 2MASS -- that is to say, that successful identification of the
RASS/BSC X-ray sources is limited only by confusion, and not by high
X-ray/optical flux ratios.

The present approach demonstrates the capability of statistical
cross-identification techniques applied using all-sky-surveys
\cite{xid}. Its advantage over previous approaches of examining
individual X-ray sources and their nearby off-band counterparts is
that we can ``Monte-Carlo'' the sky, and quantitatively state
selection efficiency for INSs.  As such, we are also able to examine
the weakness of this approach, and design a superior experiment, which
will improve on the present limit.

Improvement on the present limit on the number of INSs requires
primarily a more efficient means of statistical identification of
INSs, and not substantial numbers of new arcsec localizations of
reasonable INS candidates in the RASS/BSC.  For example, if we had
observed all 32 (spectrally soft) INS candidates with \chandra, and
there were no new INSs, this would lower the 90\% limit from 56 to 34
INSs, an improvement of a factor of $\sim$1.6.  However, if our
identification efficiency had been 80\% instead of 20\% but everything
else being the same (including having 13 ``undetermined'' sources),
our 90\% confidence upper-limit would now be $<$9 -- comparable to the
number known -- instead of $<$56, an improvement of a factor of
$\sim$5.6, which would not have required any more observations than we
undertook for the present analysis.  Thus, efficient statistical
identification strategies are more critical to INS identification than
are complete observations of ``reasonable'' candidates.  Statistical
applications such as this are becoming more common as the Virtual
Observatory\footnote{http://www.nvosdt.org} is developed \cite{nvo},
and can be expected to leverage the large all-sky databases with
modest new observations, to extract maximum science output.

\acknowledgements 

RER was supported for this work by the \chandra\ Guest Observation
program. RER acknowledges stimulating conversations with G.\ Farrar
and L.\ Bildsten on the subject of observational limits on the NS
birthrate.  We are grateful to David Kaplan, Edo Berger and Josh Bloom
for obtaining optical images in support of the present work.  We are
also grateful to the members of the \chandra\ Science Center for
production of this exquisite observatory.  This research has made use
of the NASA/ IPAC Infrared Science Archive, which is operated by the
Jet Propulsion Laboratory, California Institute of Technology, under
contract with the National Aeronautics and Space Administration.  This
publication makes use of data products from the Two Micron All Sky
Survey, which is a joint project of the University of Massachusetts
and the Infrared Processing and Analysis Center/California Institute
of Technology, funded by the National Aeronautics and Space
Administration and the National Science Foundation.  The Digitized Sky
Surveys were produced at the Space Telescope Science Institute under
U.S. Government grant NAG W-2166. The images of these surveys are
based on photographic data obtained using the Oschin Schmidt Telescope
on Palomar Mountain and the UK Schmidt Telescope. The plates were
processed into the present compressed digital form with the permission
of these institutions.  The DPOSS project was generously supported by
the Norris Foundation.

\newpage

\newpage

\clearpage
\newpage

\begin{figure}[htb]
\caption{ \label{fig:control} {\bf (a)} Distribution in Galactic
Coordinates of the 150 control fields.  These objects mimic INSs, in
that they will have no detectable off-band counterparts in the
USNO-A2, IRAS, or NVSS catalogs.  {\bf (b)} Distribution in Galactic
Coordinates of the 29 (of 150) control fields with \pnoid$>$0.90.
These largely avoid the plane, where the source density produces many
(low significance) candidate associations.  {\bf (c)} The distribution
of \pid\ values for optical sources in the 150 control fields.  The
vast majority are of low significance, as expected for spurious
associations.  However, when \approxgt few of these appear in the same
control field, as occurs in the Galactic plane, the sum of the \pid\
values removes the X-ray sources from our selection. {\bf (d)} The
cumulative distribution $N(<|b|)$) of the 29 control fields with
\pnoid$>$0.90 (solid line) compared with an equal number per sky-area
distribution (dotted line).  The observed distribution is discrepant
from the equal area distribution (the probability of producing the
observed distribution from the theoretical distribution is 3\tee{-6})
due to the effects of confusion in the Galactic plane.  }
\end{figure}

\nocite{vandokkum01}
\begin{figure}[htb]
\caption{ \label{fig:0203esiimg} R-band 5-min ESI image taken at Keck
II on 2002 Nov 5 UT, of the field of 1RXS~J0203-3438.  Cosmic rays
were removed by application of the ``L.A.COSMIC'' algorithm (van
Dokkum 2001) . North is up, east to the left. The intensity gradient
across the field is due to a bright optical source which is
1.5\arcmin\ off-image.  The localization of the \chandra\ X-ray source
is indicated by the cross, with a 0.6\arcsec\ (2\sig) positional
uncertainty. For this figure, relative astrometry was performed with a
second X-ray source detected in the HRC image, which is coincident
with a bright DSS source, and is also detected as a point source
(0.8\arcsec\ seeing) in the ESI image. The \chandra\ source is
spatially associated with an optical point source, with two
nebulosities to the north-east and to the north.  }
\end{figure}

\begin{figure}[htb]
\caption{ \label{fig:1328} Field of 1RXS~J1328$-$3554, from UK Schmidt
48-inch survey (left panel, Epoch 1975.27), second epoch survey UKSTU
Schmidt survey (center panel, Epoch 1991.2), and 2MASS $J$-band image
(right panel, Epoch 1999.45, scaled logarithmically in intensity).
Hash marks denote the localization of the \chandra\ X-ray source.
Circles mark the positions of the 2MASS IR sources found in the field.
The epoch of the \chandra\ observation is 2001.2.  The \chandra\
source is coincident with a red, apparent high proper motion object
($\sim$165 mas/yr), which we interpret as a nearby low-mass star. }
\end{figure}

\begin{figure}[htb]
\caption{ \label{fig:1450} DPOSS g-band image of the field of
1RXS~J1450+6559.  The large circle is the 3$\sigma$ RASS localization.
The small (black) circle is the 1\arcsec\ \chandra\ positional
uncertainty.  The optical source coincident with the \chandra\
position is fainter than the DSS plate limit, and so not included in
the USNO-A2 catalog.  }
\end{figure}

\begin{figure}[htb]
\caption{ \label{fig:lris} KECK/LRIS optical spectra (not flux
calibrated) of proposed optical counterparts.  {\bf Top Panel}: 1RXS
J1450+6559 contains three prominent emission lines at 6555\AA,
6668\AA, and 7055\AA. Absorption features at 6862, 7180, 7234, 7596,
and 7621\AA\ are sky lines.  {\bf
Center Panel}: 1RXS J1639+5656 shows one prominent broad emission
line, at 7412\AA.  The absorption features (in particular, those at
7233, 7267\AA\ and 7595, 7625\AA) are due to imperfect sky subtraction. {\bf
Bottom Panel:} Sky spectrum.  }
\end{figure}

\begin{figure}[htb]
\caption{ \label{fig:1639} DPOSS g-band image of field
1RXS~J1639+5656.  The large circle is the 42\arcsec\ (3$\sigma$)
RASS/BSC localization uncertainty; the small (black) circle, indicated
by the hash marks, is the 1\arcsec\ radius \chandra\ localization,
which is positionally coincident with an extended source, an apparent
AGN at $z=1.65$.  }
\end{figure}

\begin{figure}[htb]
\caption{ \label{fig:dss} DSS fields of the 8 candidate INSs observed
with \chandra.  The large circle is the 3$\sigma$ RASS/BSC positional
uncertainty, labeled with the RASS source name.  Those with X-ray
sources detected with \chandra\ have a cross, representing the
1\arcsec\ \chandra\ positional uncertainty, labeled with the \chandra\
source name.  In the upper-right corner, we give the Galactic
coordinates (deg) of the X-ray source, and the equivalent Hydrogen
column density (\nh) in units of \perval{cm}{-2}. In the lower-left
corner, we list \pid, the probability the RASS/BSC source is
associated with one of the off-band sources; USNO-A2 sources which
individually have a non-zero probability of being associated with the
RASS/BSC source are labeled with their individual \pid\ value.  We
also list in the lower-left corner the RASS/BSC hardness ratio (HR1)
and PSPC countrate.  (see more discussion in text).  }
\end{figure}

\begin{figure}[htb]
\caption{ \label{fig:pxid} Probability distribution $P_{\rm XID}(p)$,
where $p$ is the probability that an INS in our survey field is found
by our statistical-identification procedure. 
}
\end{figure}

\begin{figure}[htb]
\caption{ \label{fig:pbg} The probability distribution $P_{\rm BG}$
(Eq.~\ref{eq:comb}) that there are $N$ INSs in our selection of 32,
yet end up with only 2 INSs in the 15 following the removal of 17 from
our sample due to associated optical sources in the field.
Here, $N_{\rm INS, min}=2$, $T=32$, $BG=17$. 
}
\end{figure}

\clearpage
\pagestyle{empty}
\begin{figure}[htb]
\PSbox{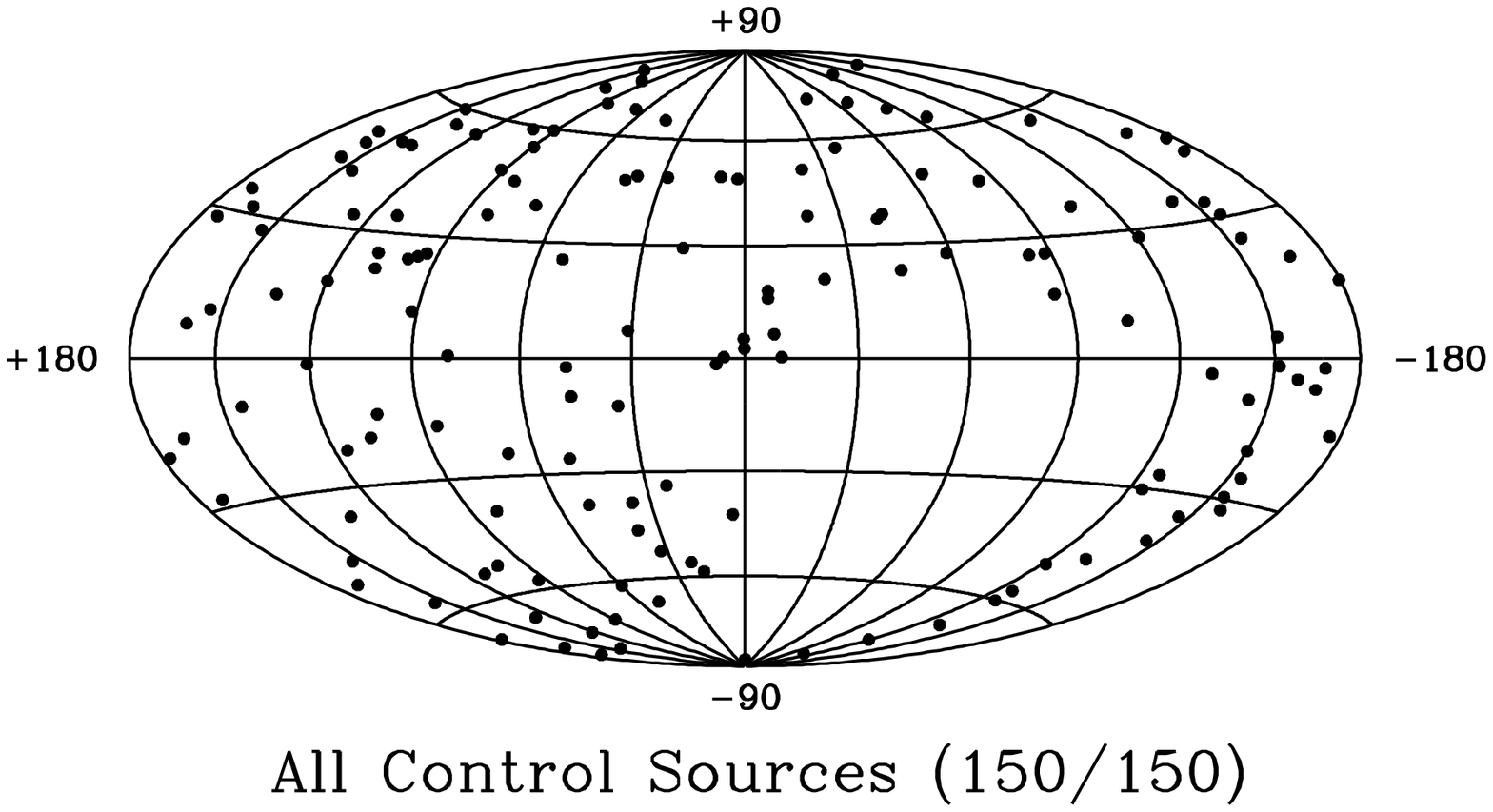 hoffset=-80 voffset=-80}{14.7cm}{21.5cm}
\FigNum{\ref{fig:control}a}
\end{figure}

\clearpage
\pagestyle{empty}
\begin{figure}[htb]
\PSbox{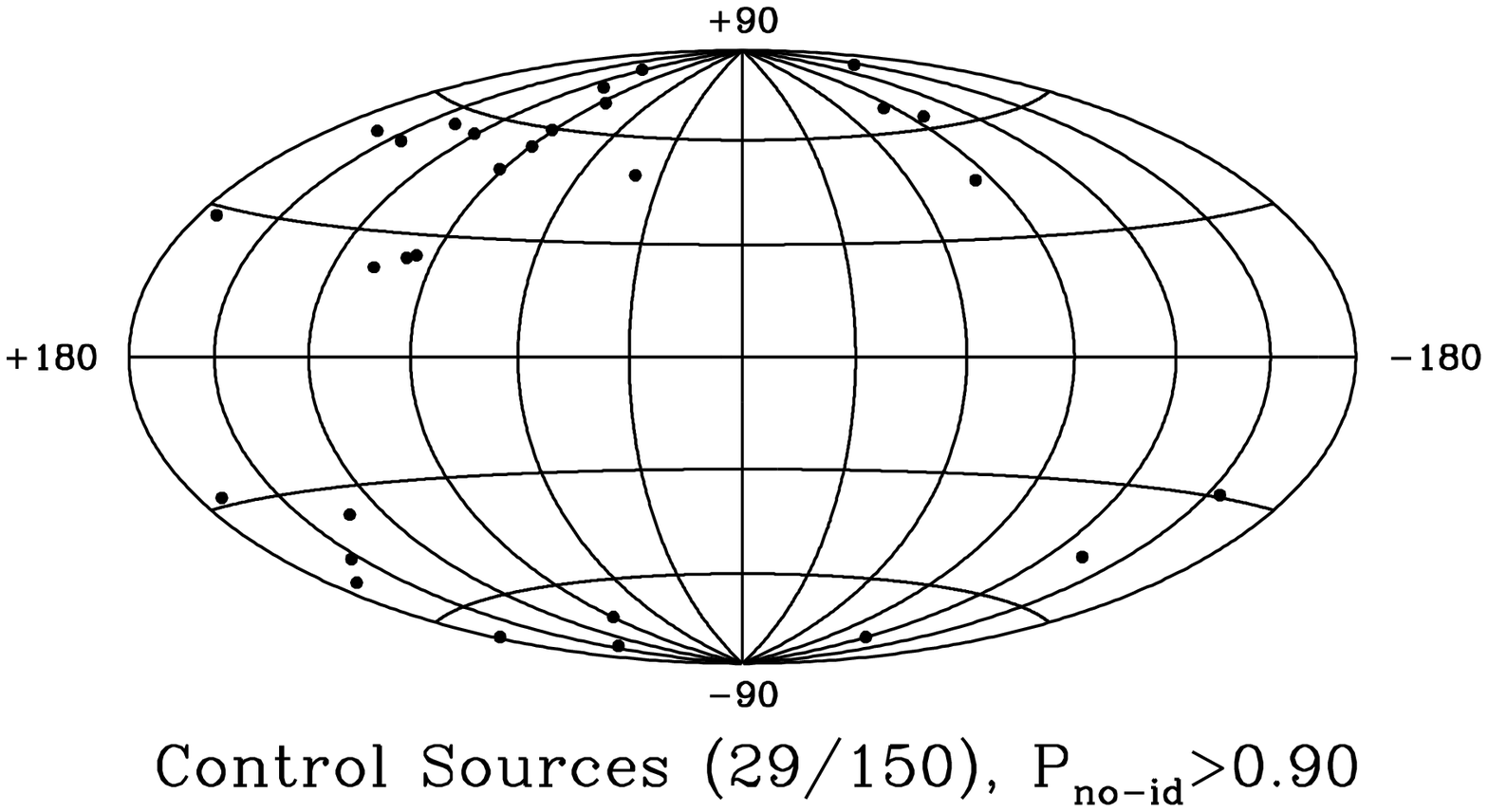 hoffset=-80 voffset=-80}{14.7cm}{21.5cm}
\FigNum{\ref{fig:control}b}
\end{figure}

\clearpage
\pagestyle{empty}
\begin{figure}[htb]
\PSbox{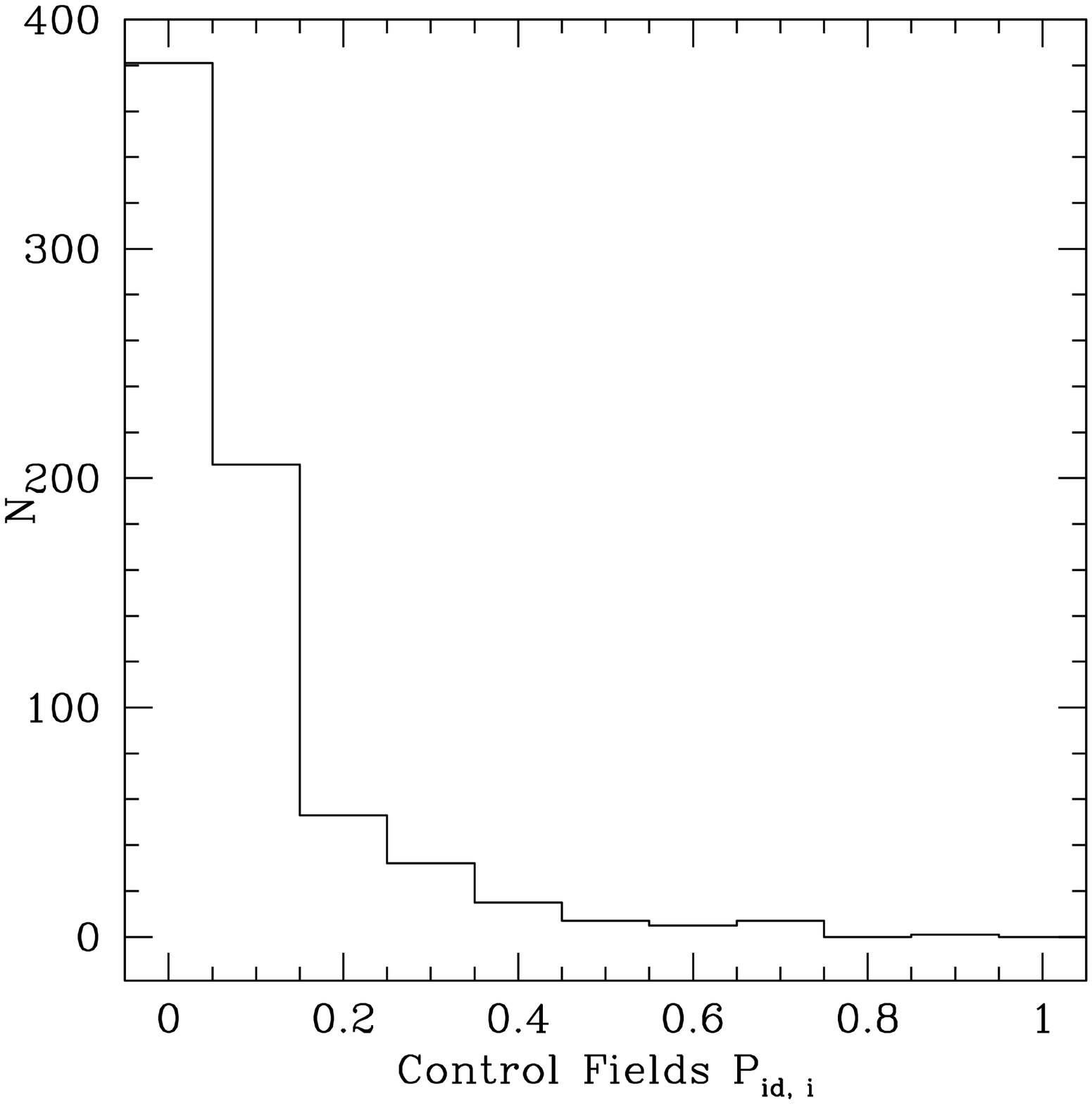 hoffset=-80 voffset=-80}{14.7cm}{21.5cm}
\FigNum{\ref{fig:control}c}
\end{figure}

\clearpage
\pagestyle{empty}
\begin{figure}[htb]
\PSbox{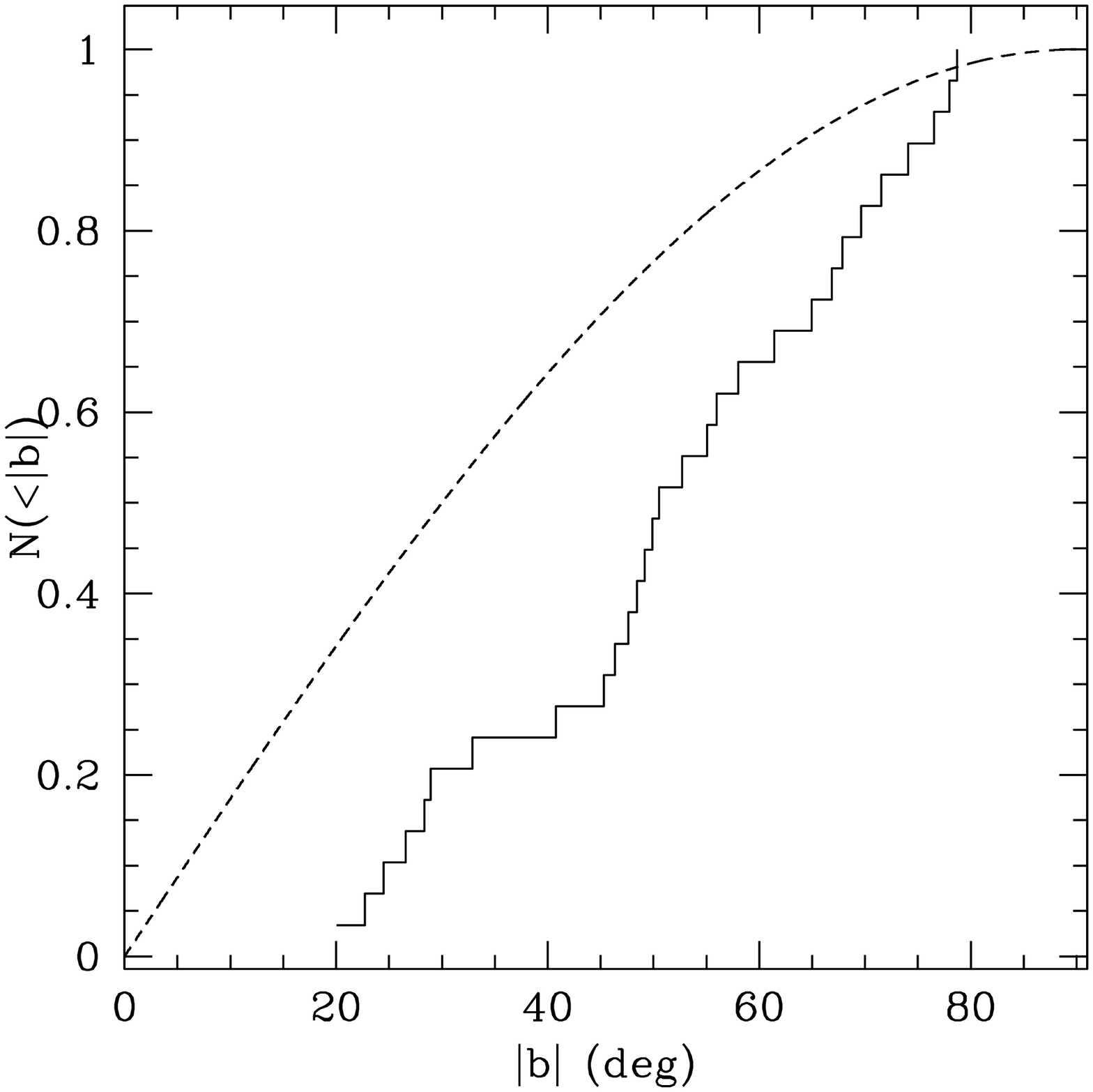 hoffset=-80 voffset=-80}{14.7cm}{21.5cm}
\FigNum{\ref{fig:control}d}
\end{figure}

\clearpage
\pagestyle{empty}
\begin{figure}[htb]
\PSbox{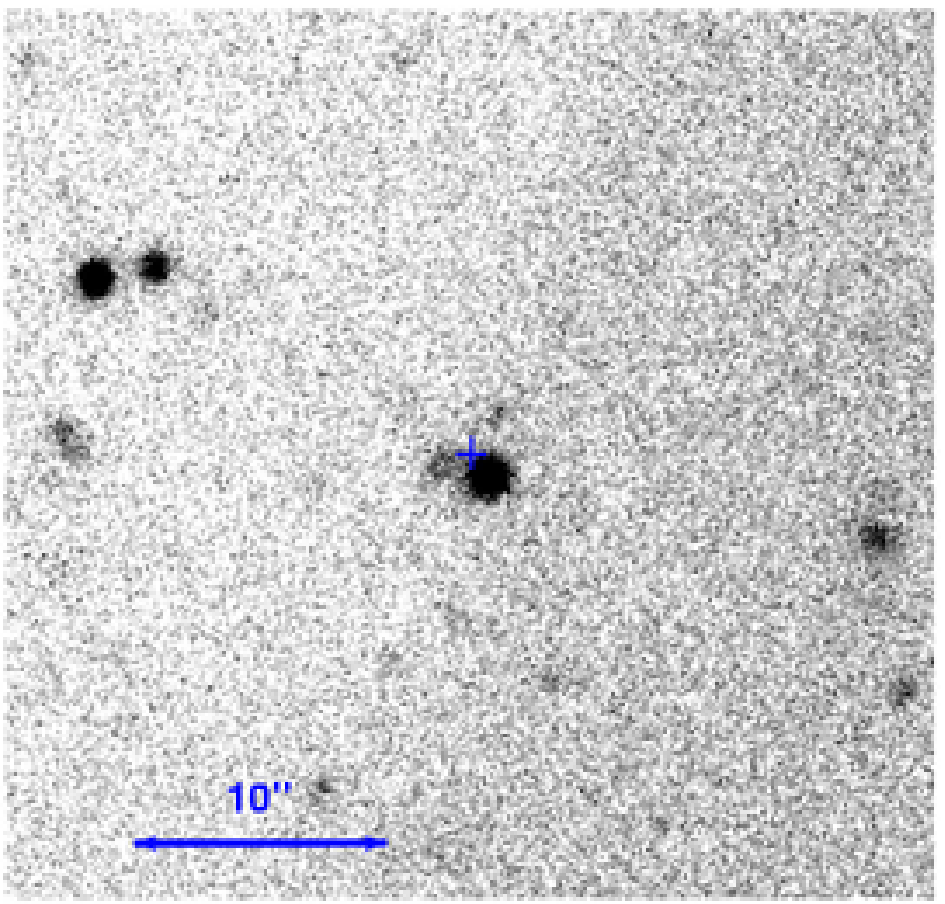  hoffset=+120 voffset=+180}{14.7cm}{21.5cm}
\FigNum{\ref{fig:0203esiimg}}
\end{figure}

\clearpage
\pagestyle{empty}
\begin{figure}[htb]
\PSbox{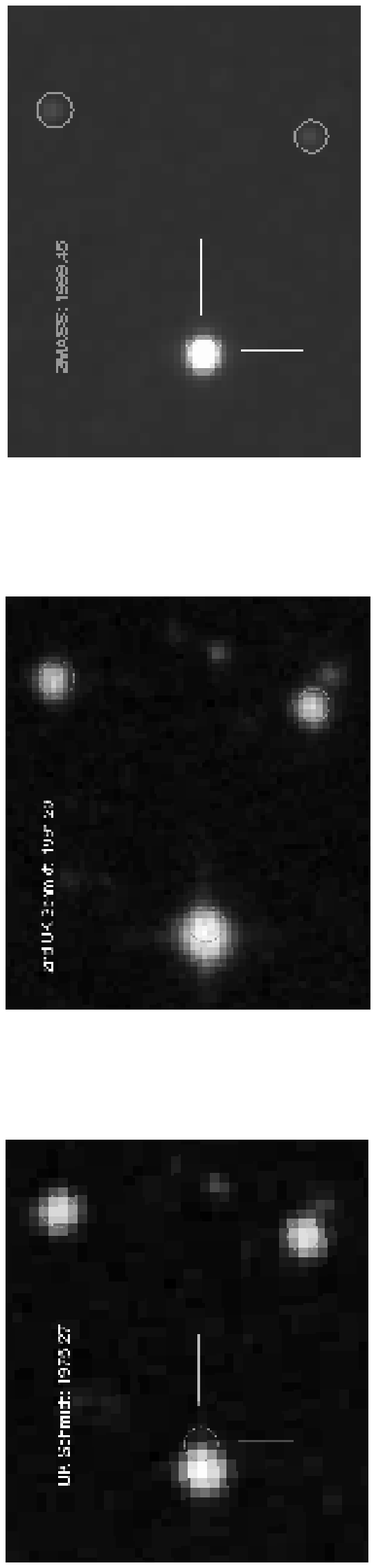  hoffset=-80 voffset=-80}{14.7cm}{21.5cm}
\FigNum{\ref{fig:1328}}
\end{figure}

\clearpage
\pagestyle{empty}
\begin{figure}[htb]
\PSbox{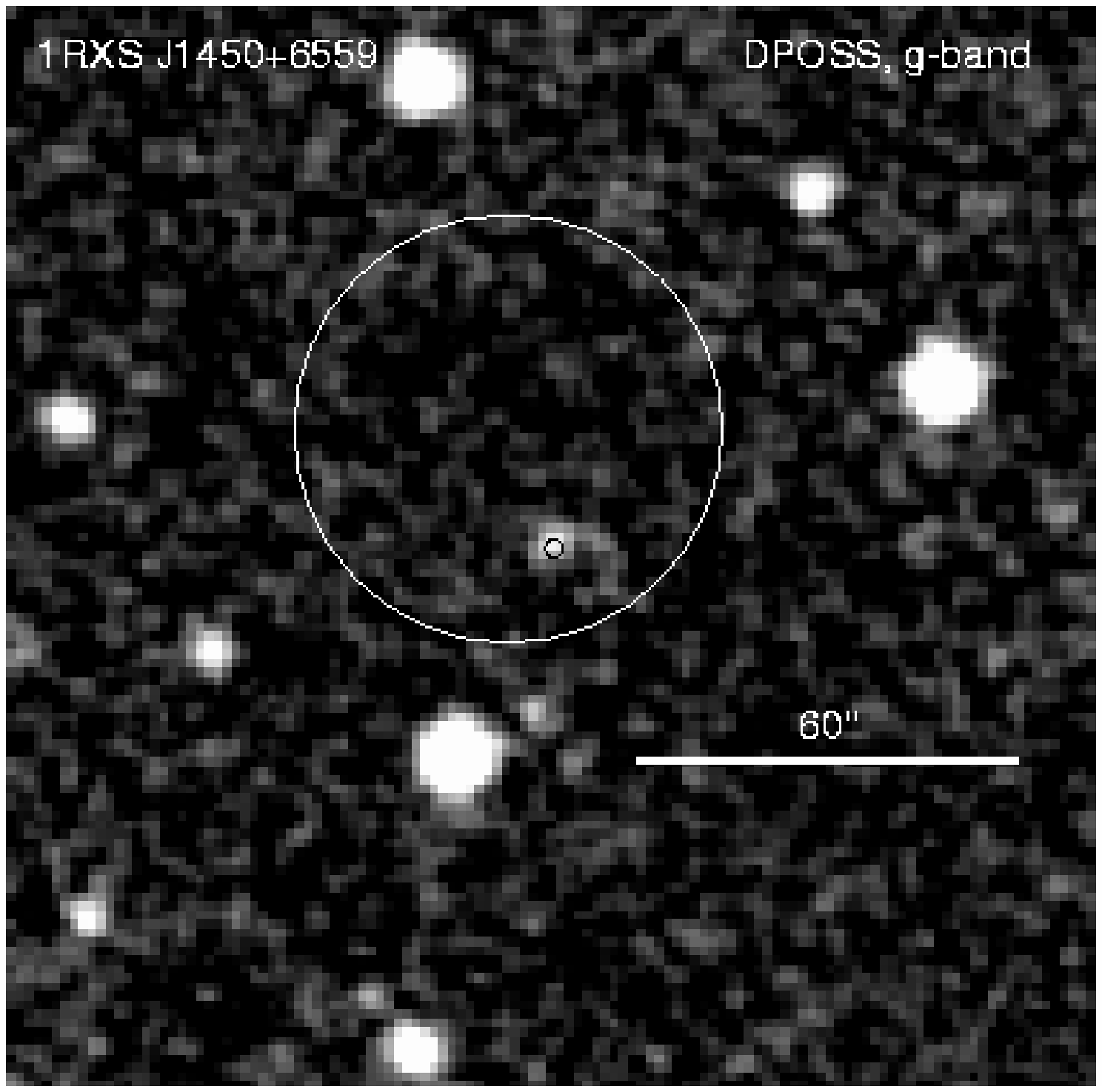   hoffset=-80 voffset=-80}{14.7cm}{21.5cm}
\FigNum{\ref{fig:1450}}
\end{figure}

\clearpage
\pagestyle{empty}
\begin{figure}[htb]
\PSbox{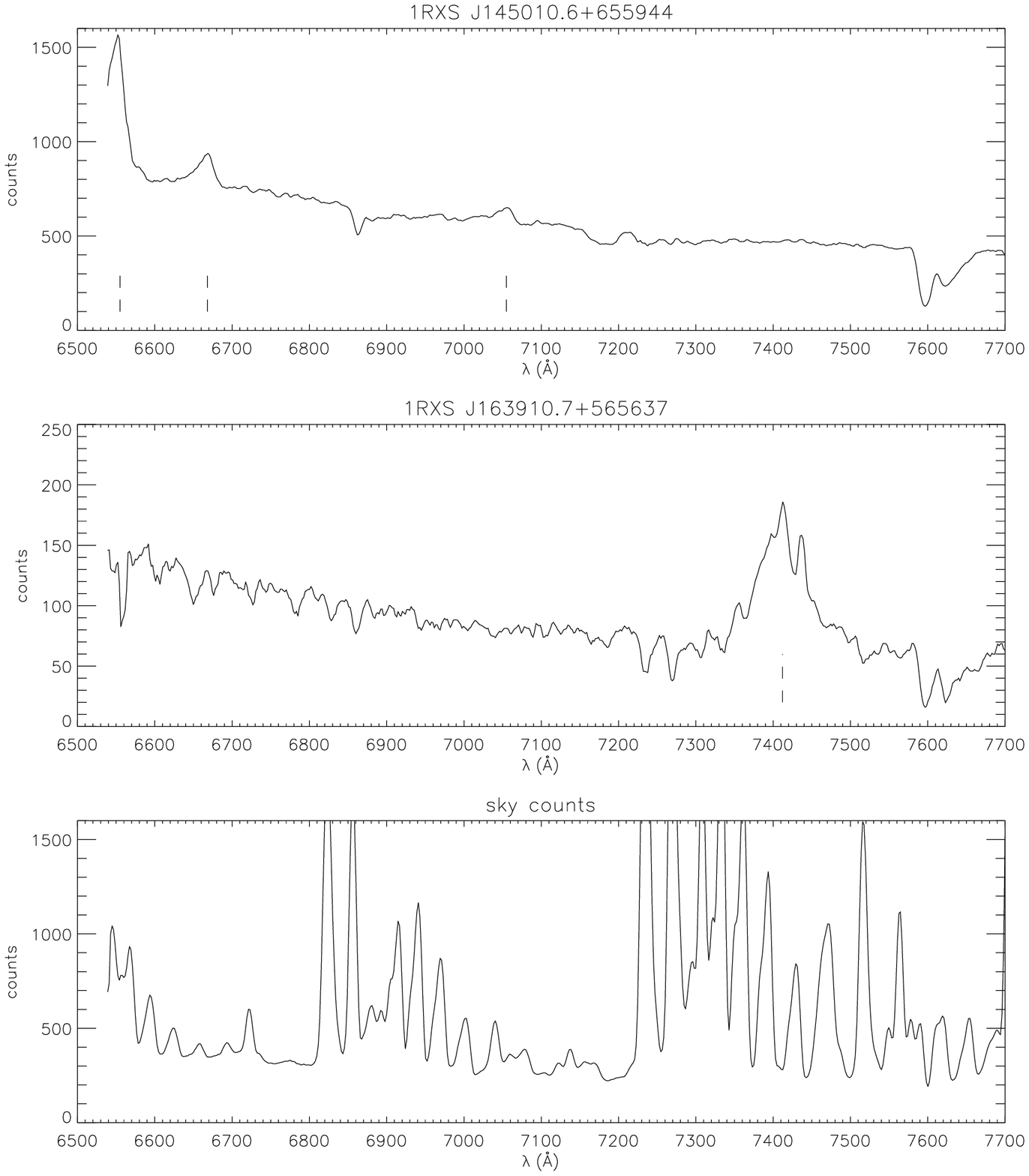 hoffset=-80 voffset=-80}{14.7cm}{21.5cm}
\FigNum{\ref{fig:lris}}
\end{figure}

\clearpage
\pagestyle{empty}
\begin{figure}[htb]
\PSbox{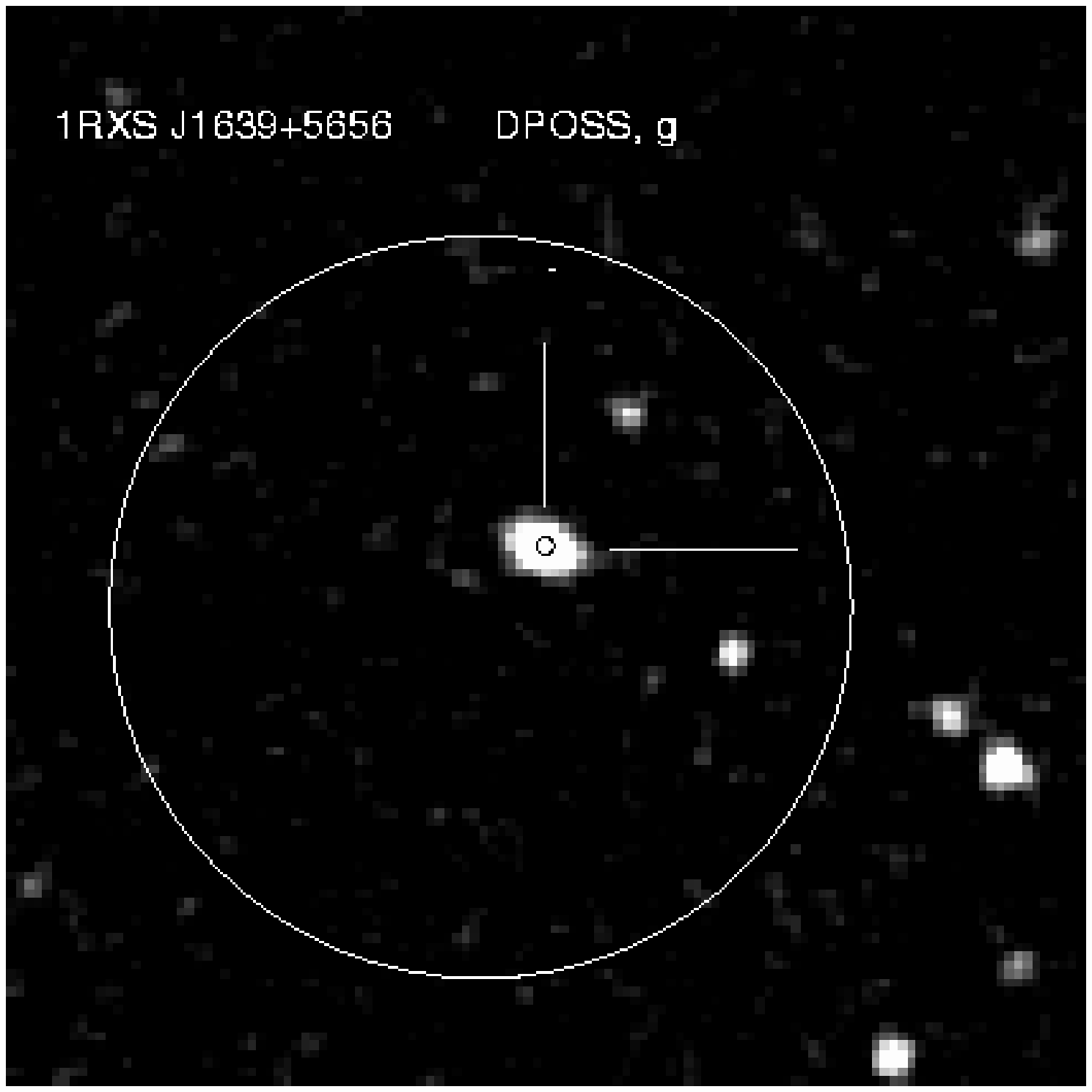 hoffset=-80 voffset=-80}{14.7cm}{21.5cm}
\FigNum{\ref{fig:1639}}
\end{figure}

\clearpage
\pagestyle{empty}
\begin{figure}[htb]
\PSbox{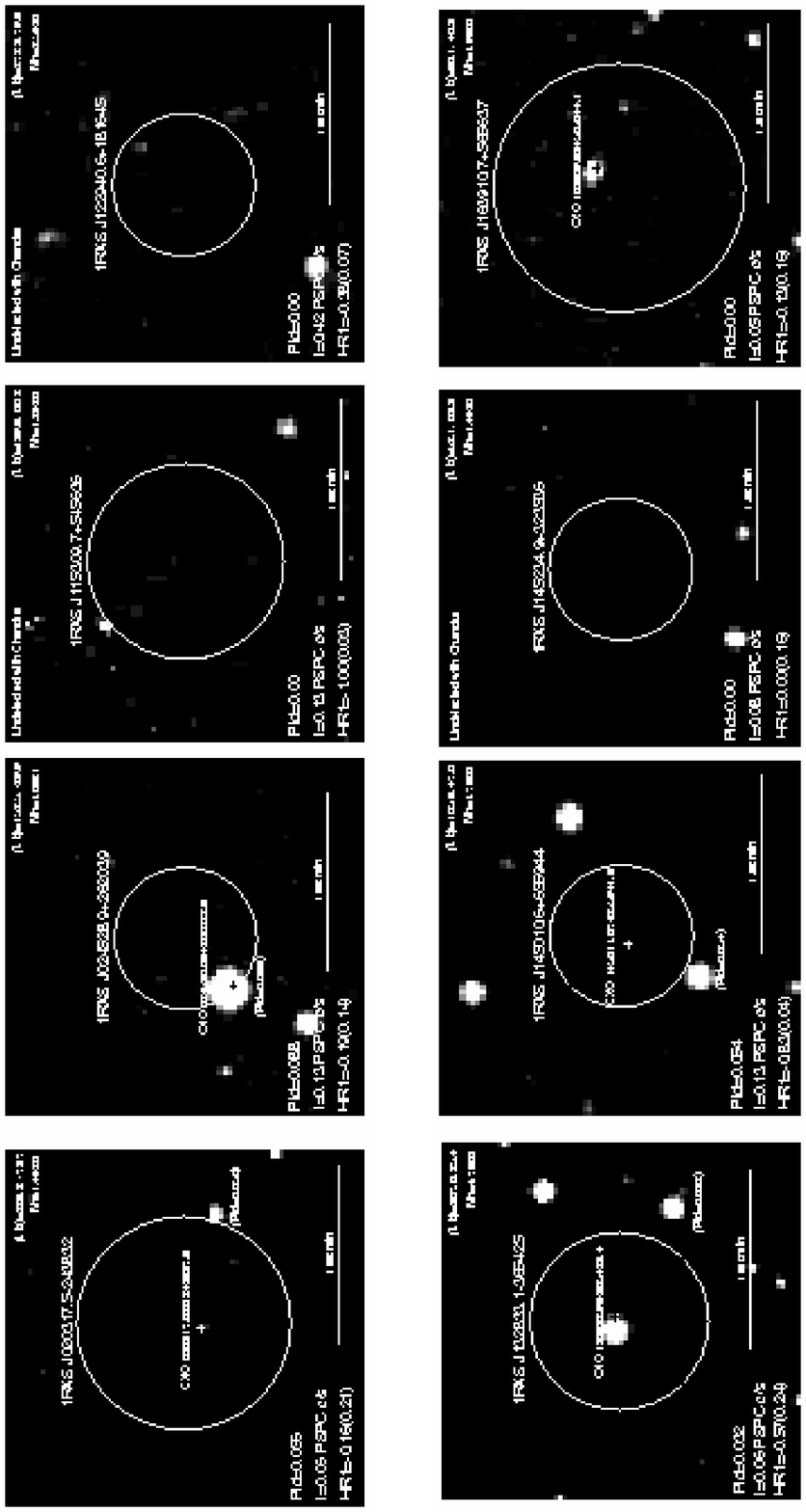  hoffset=-80 voffset=-80}{14.7cm}{21.5cm}
\FigNum{\ref{fig:dss}}
\end{figure}

\clearpage
\pagestyle{empty}
\begin{figure}[htb]
\PSbox{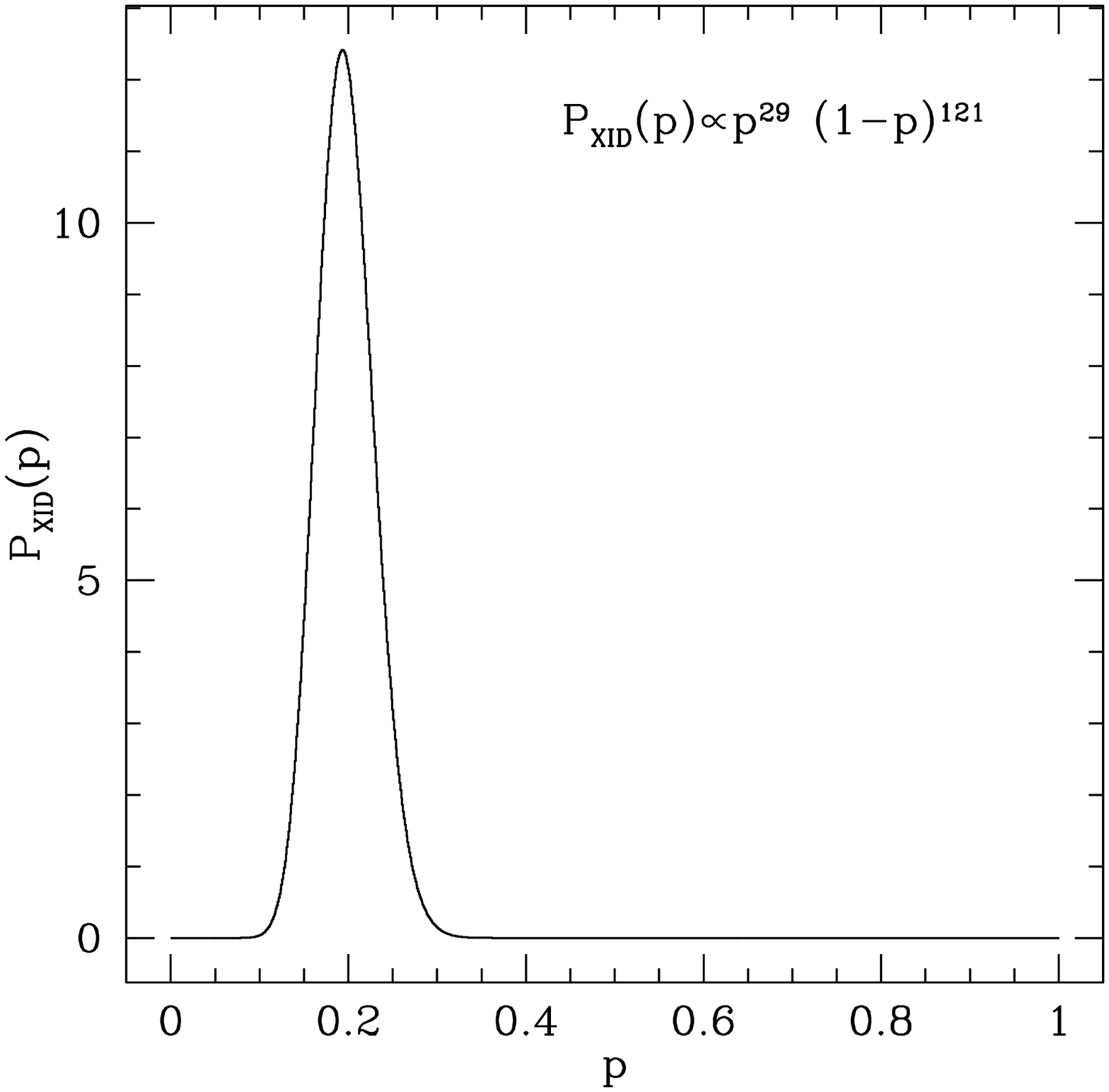  hoffset=-80 voffset=-80}{14.7cm}{21.5cm}
\FigNum{\ref{fig:pxid}}
\end{figure}

\clearpage
\pagestyle{empty}
\begin{figure}[htb]
\PSbox{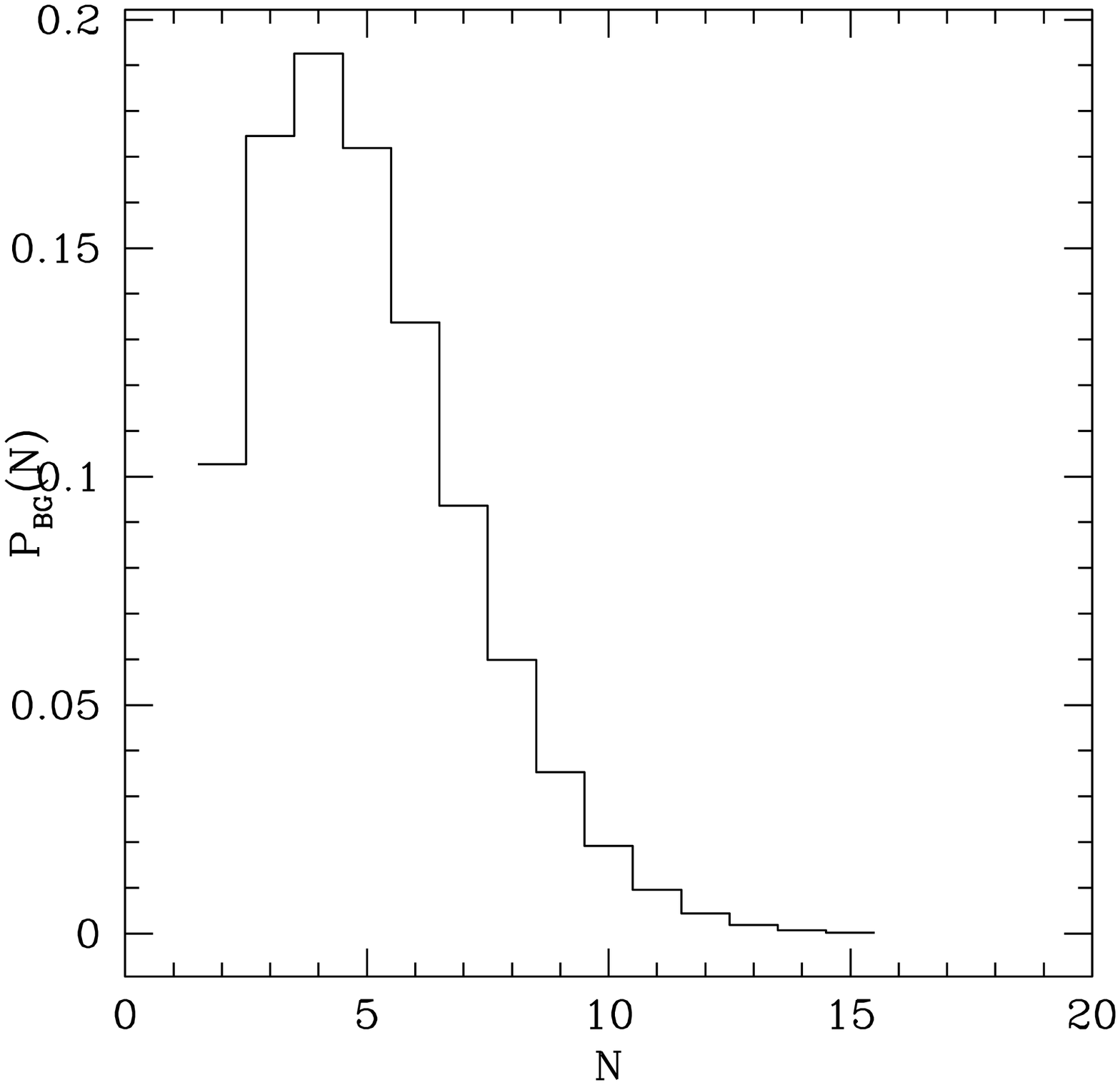  hoffset=-80 voffset=-80}{14.7cm}{21.5cm}
\FigNum{\ref{fig:pbg}}
\end{figure}

\clearpage
\newpage

\nocite{treves00}
\begin{deluxetable}{lll}
\tablecaption{Identified  INSs \label{tab:ins}}
\tablehead{
\colhead{INS} &
\colhead{1$-$\pnoid} & 
\colhead{High \pid\ Object $^a$}
}
\startdata
MS 0317.7$-$6647	& n/a	&  n/a			\\
RX J185635$-$3754	&  0.89	& $B$=17.4; $r$=6.7\arcsec  	\\
RX J0720.4$-$3125	& 1.0 & $B$=20.2, $r=$6.4\arcsec 	 \\
RX J0420.0$-$522  	& n/a	&	n/a			\\
RX J1308.6+2127		& 0.0	&	no sources		\\
RX J1605.3+3249		& 0.0	&  	no sources		\\
RX J0806.4$-$4132	& n/a	& n/a		\\ 
RX J2143.0+0654 & 0.89  & $B=19.4$, $r=22.4$\arcsec	\\
\enddata
\tablecomments{X-ray sources classified as INSs (Treves \etal 2000;
\citenp{zampieri01}) n/a = Not included in our analysis (either at
$\delta<-39$ or not in RASS/BSC).\\ $^a$ The single object which has
the highest probability of association.}
\end{deluxetable}

\newcommand{\notdefinitive}{undetermined}
\newcommand{\DEFINITIVE}{not an INS}
\newcommand{\CHANDRA}{\chandra}
\newcommand{\ROSOBS}{\rosat/HRI}
\newcommand{\uok}{}
\newcommand{\notid}{\nodata}
\newcommand{\uprob}{}

\begin{deluxetable}{llllccl}
\scriptsize
\tablecaption{Table of Thirty Two INS Candidates\label{tab:inscandidates}}
\tablehead{
\colhead{1RXS~J} 	& 
\colhead{RBS \#} 	&
\colhead{type} 		& 
\colhead{Name }		& 
\colhead{\chandra\ /HRI Obs.?}	& 
\colhead{INS?}		&
\colhead{ID. Ref. }	\\
}
\startdata
020146.5+011717 &     269  	&  *		&\nodata& \nodata	&\uok   \notdefinitive\  &       	\\
020317.5$-$243832 &		&  AGN		&CXO J020317.62$-$243837.8& \CHANDRA	&\uok   \DEFINITIVE 	  &   (\S~\ref{sec:chandra})	\\
024946.0$-$382540 &		&  \notid	&\nodata& \nodata	&\uprob \notdefinitive\  &       	\\
024528.9+262039 &		&  *		&\nodata& \CHANDRA	&\uok   \DEFINITIVE 	  &   (\S~\ref{sec:chandra})	\\
031413.7$-$223533 &		& nova/star     &EF Eri & \nodata	&\uok   \DEFINITIVE 	  & 1,2   	\\
032620.8+113106 &		& T-Tauri	&\nodata& \nodata	&\uprob \notdefinitive   & 3     	\\
041215.8+644407 &		&  Fl* 	&G 247-15	& \nodata	&\uprob \notdefinitive   & 10    	\\
043334.8+204437 &		&  Fl* 		&G 8-41	& \nodata	&\uprob \notdefinitive   & 10    	\\
051541.7+010528 &		&CV/AM-Her&V* V1309 Ori	& \nodata	&\uok   \DEFINITIVE  	  &  4    	\\
051723.3$-$352152 &		&  MV:e	&EUVE J0517-35.3& \nodata	&\uprob \notdefinitive   &       	\\ 
075556.7+832310 &		& Fl* 	&EUVE J0755+83.3& \nodata	&\uok   \notdefinitive   & 10    	\\
091010.2+481317 &		& Sy1 & QSO B0906+484	& \nodata	&\uok   \notdefinitive   &       	\\ 
094432.8+573544 &		& AGN &  DPOSS 094432.42+573534.9& \ROSOBS	&\uok   \DEFINITIVE 	  &   (\S~\ref{sec:rosat})   	\\ 
104710.3+633522 &		& CV/DQ Her&V* FH Uma	& \nodata	&\uok   \DEFINITIVE  	  & 5     	\\
115309.7+545636 &		& \notid 	&\nodata& \CHANDRA	&\uok   \DEFINITIVE 	  &   (\S~\ref{sec:chandra}) 	\\
122940.6+181645 &     1116 	& BL-Lac 	&\nodata& \CHANDRA	&\uok   \DEFINITIVE 	  &  (\S~\ref{sec:chandra})  	\\
123319.0+090110 &		& **		&GJ 473B& \nodata	&\uok   \DEFINITIVE 	  & 6     	\\
125015.2+192357 &		& Sy1		&\nodata& \nodata	&\uok   \notdefinitive   &       	\\
130034.2+054111 &		& Fl* & V* FN Vir   	& \nodata	&\uprob \notdefinitive   &       	\\
130402.8+353316 &		& QSO& QSO B1301+358	& \nodata	&\uok   \notdefinitive   &       	\\
130547.2+641252 &		&  \notid	&	& \ROSOBS	&\uok   \DEFINITIVE 	  &   (\S~\ref{sec:rosat})   	\\ 
130753.6+535137 &     1219	& CV/AM-Her&V* EV UMa	& \ROSOBS	&\uok   \DEFINITIVE 	  &  (\S~\ref{sec:rosat})    	\\ 
130848.6+212708 &     1223	& INS  &		& \nodata	&\uok   INS 		  & 7     	\\
132833.1$-$365425 & 		& HiPM*		&CXO  J132832.98$-$365423.4	& \CHANDRA	&\uprob \DEFINITIVE 	  &   (\S~\ref{sec:chandra}) 	\\
134210.2+282250 &     1306 	& CV in M3	&\nodata& \nodata	&\uprob \DEFINITIVE 	  & 8     	\\
145010.6+655944 & 		&  CV		&\nodata& \CHANDRA	&\uok   \DEFINITIVE 	  &    (\S~\ref{sec:chandra})	\\
145234.9+323536 & 		&  \notid	&\nodata& \CHANDRA	&\uok   \DEFINITIVE 	  &   (\S~\ref{sec:chandra}) 	\\
160518.8+324907 &     1556	& INS  		&\nodata& \nodata	&\uok   INS 		  & 9     	\\
163421.2+570933 & 		&  HiPM**	&2MASS~J163420.44+570944.0	& \ROSOBS	&\uprob \DEFINITIVE 	  & (\S~\ref{sec:rosat})   	\\ 
163910.7+565637 & 		&  AGN 		&CXO J163909.83+565644.1	& \CHANDRA	&\uprob \DEFINITIVE 	  & (\S~\ref{sec:chandra})    	\\  
231543.7$-$122159 &     1970	& HiPM*	&L~863-30	& \nodata	&\uok   \notdefinitive   &       		\\
231728.9+193651 &     1978	& Fl*		& G~68-5& \nodata	&\uok   \notdefinitive   &       	\\ \hline
\enddata
\tablecomments{
\uprob=bright ($V<$15) source in DSS, but not cataloged in USNO-A2,
within 30\arcsec\ of RASS/BSC position.
\DEFINITIVE=Definitively not an INS. 
INS= Previously identified as an INS. *=star; Fl*=flare star; HiPM*=high proper-motion star; HiPM**=high proper-motion binary; 
\notdefinitive=X-ray source which has not been definitively excluded as an INS. 
Refs: 1,  \acite{watson87};  
2, \acite{beuermann91}; 
3, \acite{li00}; 
4, \acite{demartino98}; 
5, \acite{singh95}; 
6, \acite{marino00}; 
7, \acite{hambaryan02}; 
8, \acite{dotani99}; 
9, \acite{motch99}; 
10, this work. 
}
\end{deluxetable}

\begin{deluxetable}{lllrrrlr} 
\scriptsize
\tablecaption{\label{tab:targets} INSs Candidate Source and \chandra\ Observation List}
\tablehead{
\colhead{} & 
\colhead{} & 
\colhead{PSPC} & 
\colhead{} & 
\colhead{RASS 1$\sigma$} &
\colhead{\chandra} &
\colhead{Obs. Start} &
\colhead{Dur.}   \\
\colhead{1RXS~J} & 
\colhead{\pnoid} & 
\colhead{c/s} & 
\colhead{HR1 (\ppm)$^a$} & 
\colhead{(arcsec)} &
\colhead{ObsID} & 
\colhead{(TT)} &
\colhead{(sec)}   
}
\startdata
0203$-$2438  &0.94 & 0.06     & -0.16 (0.21) &  12	& 1973& 2001 Jan 28 11:19&    910             	\\
0245+2620  &0.91 & 0.13     & -0.19 (0.14) &  8 	& 1974& 2001 Jan 16 03:06&    821             	\\
1153+5456  &1.0	 & 0.13     & -1.00 (0.03) &  11	& 1975& 2001 Nov 19 22:54&  846               	\\
1229+1816  &1.0	 & 0.42     & -0.38 (0.07) &  8		& 1976& 2001 Mar 24 04:46&   1124             	\\
1328$-$3654  &0.96 & 0.06     & -0.57 (0.24) &  10	& 1977& 2001 Mar 24 06:40&   1325             	\\ 
1450+6559  &0.95 & 0.13     & -0.83 (0.04) &  8		& 1978& 2001 Sep 07 17:33&  823               	\\
1452+3235  &1.0	 & 0.08     & 0.00  (0.16) &  8		& 1979& 2001 Mar 24 05:25&    1334            	\\
1639+5656  &1.0	 & 0.05     & -0.13 (0.16) &  14	& 1980& 2001 Sep 07 18:10& 1172               	\\  
\enddata
\tablecomments{$^a$. HR1 is a hardness ratio (see text), for which HR1=0.0 corresponds to 200 eV, and 
lower values are $<$200 eV.  } 
\end{deluxetable}

\begin{deluxetable}{lrlll}
\scriptsize
\tablecaption{\label{tab:sources} \chandra\ and ROSAT/HRI X-ray Source Localizations and Classifications}
\tablehead{
\colhead{1RXS J} & 
\colhead{predicted I$^a$} & 
\colhead{observed I} & 
\colhead{ Source Position} &
\colhead{}   \\
\colhead{} & 
\colhead{(c/ksec)} & 
\colhead{(c/ksec)} & 
\colhead{R.A./dec. (J2000)} & 
\colhead{Notes} 
}
\startdata
 0203$-$2438 & 70    & 35(6) & CXO J020317.626$-$243837.8		& AGN? \\
 0245+2620 & 140   & 41(7) & CXO J024530.08+262022.8	        & M3-star\\
 1153+5456 & 30    &$\leq$4.7 & undetected					& n/a	\\
 1229+1816& 400   & $\leq$2.7   & undetected					& n/a	\\
 1328$-$3654& 50    &687(25)& CXO  J132832.98$-$365423.4      	& proper-motion star  \\
 1450+6559& 50    & 62(8) & CXO J145011.07+655941.8		& LRIS spectrum; CV\\
 1452+3235& 100   & $\leq$2.3   & undetected					&  n/a 		\\
 1639+5656& 50    & 47(6) & CXO 163909.83+565644.1	& extended, LRIS specturm; AGN \\  \hline
 0944+5735& 45	  & 9(1) & 1RXH~J094431.8+573538		& AGN? 			\\
 1305+6412& 55	  & $<$3&  undetected			& n/a			\\
 1307+5351& 600	  & $<$0.5& undetected			& CV	\\ 
 1634+5709& 55	  & 38(4)& 1RXH~J163421.2+570941		& High proper-motion binary \\
\enddata
\tablecomments{$^a$. Predicted countrate, based on spectral hardness and \rosat/PSPC countrate.  Count-rate upper limits are not formally detection
limits, but limits on the average countrate.  Nominal positional
uncertainties for blind pointing of \chandra\ are 1\arcsec. The
positional uncertainty for CXO~J020317.626$-$243837.8 is
0.4\arcsec\ due to a second X-ray source detected in the HRC image. }
\end{deluxetable}

\begin{deluxetable}{lrrr}
\scriptsize
\tablecaption{\label{tab:cat}Off-band Catalog Detections at
$\sim$arcsec Localizations}
\tablehead{
\colhead{} &
\colhead{2MASS} & 
\colhead{2MASS} & 
\colhead{} \\
\colhead{CXO} &
\colhead{(quicklook)} & 
\colhead{(catalog)} & 
\colhead{DPOSS} 
}
\startdata
CXO~020317.60$-$243839.5	& no		& \nodata				& \nodata			\\
CXO~ 024530.08+262022.8	& yes		& 0245300+262023			& 024530.10+262021.3		\\
			&		& $(J,H,K)$=9.45(3),8.73(3),8.58(3)	& $(g,r,i)$=14.24,13.54,sat.	\\
CXO~132832.98$-$365423.4   & yes	& \nodata				& \nodata			\\
CXO~ 145011.07+655941.8	& no		& \nodata				& 145011.13+655941.7		\\
			&		&					& $(g,r,i)$=19.49,20.09,19.00   \\
CXO~163909.83+565644.1	& yes		& 1639099+565644			& 163909.88+565643.6		\\
			&		& $(J,H,K)$=15.77(8),15.2(1),14.5(1)    & $(g,r,i)$=18.37,17.92,17.47   \\	\hline
1RXH~J094431.8+573538   &no		& \nodata\				& 094432.42+573534.9		\\
			&		&					& $(g,r,i)$=19.68,20.59.20.11	\\
1RXH~J163421.2+570941	& yes		& 163420.44+570944.0			& \nodata			\\
			&		& $(J,H,K)=8.50(1), 8.04(2), 7.77(2)$		& \nodata\ 			\\

\enddata
\tablecomments{ sat.=saturated. Numbers in parenthesis are the uncertainty in the preceeding digits. }
\end{deluxetable}

\end{document}